\documentclass[12pt]{article}

\usepackage{epsfig}

\begin{document}

% commands to simplify typing
\renewcommand{\d}{\partial}
\newcommand{\X}{{\cal X}}
\renewcommand{\L}{{\cal L}}
\newcommand{\A}{{\cal A}}
\newcommand{\B}{{\cal B}}
\newcommand{\C}{{\cal C}}
\newcommand{\E}{{\cal E}}

\newcommand{\R}{R}

\newcommand{\Ltilde}{\tilde{{\cal L}}}
\newcommand{\Mtilde}{\tilde{M}}
\newcommand{\ntilde}{\tilde{n}}
\newcommand{\Qtilde}{\tilde{Q}_{n,\tilde{n}}}
\newcommand{\qbar}{\bar{q}}
\newcommand{\psibar}{\bar{\psi}}
\newcommand{\dom}{\mbox{dom}}

\title{{\bf Nonequivalent Statistical Equilibrium Ensembles and
       Refined Stability Theorems for Most Probable Flows}}
\author{Richard S. Ellis,  Kyle Haven, and  Bruce Turkington
\\  Department of Mathematics and Statistics \\
University of Massachusetts\\ Amherst, MA 01003}

\maketitle

\begin{abstract}
\small
\noindent 
Statistical equilibrium models of coherent structures in
two-dimensional and barotropic quasi-geostrophic turbulence 
are formulated using canonical and microcanonical ensembles, 
and the equivalence or nonequivalence of ensembles is 
investigated for these models.  The main results show that 
models in which the global invariants are treated 
microcanonically give richer families of equilibria than 
models in which they are treated canonically.  Such global 
invariants are those conserved quantities for ideal dynamics
which depend on the large scales of the motion; they include 
the total energy and circulation.  For each model a variational 
principle that characterizes its equilibrium states is derived 
by invoking large deviations techniques to evaluate the 
continuum limit of the probabilistic lattice model.  An 
analysis of the two different variational principles resulting 
from the canonical and microcanonical ensembles reveals that 
their equilibrium states coincide only when the microcanonical 
entropy function is concave.  These variational principles also 
furnish Lyapunov functionals from which the nonlinear stability 
of the mean flows can be deduced.  While in the canonical model 
the well-known Arnold stability theorems are reproduced, in the
microcanonical model more refined theorems are obtained which
extend known stability criteria when the microcanonical and 
canonical ensembles are not equivalent.  A numerical example 
pertaining to geostrophic turbulence over topography in a zonal 
channel is included to illustrate the general results.

\vspace{.2in}

\end{abstract}

\normalsize
%==========================================================================
\section{Introduction}

\noindent
A prominent feature of two-dimensional turbulence is the formation of
large-scale coherent structures among the small-scale fluctuations of
the vorticity field \cite{McWilliams1,SBL}.  This self-organization
behavior results from the conservation of both energy and enstrophy
(the spatial second moment of vorticity) in inviscid, incompressible
two-dimensional flow, which causes a net flux of energy toward large
scales and a net flux of enstrophy toward small scales
\cite{Kraichnan}.  As a consequence, a freely-evolving flow gradually
tends toward an equilibrium state consisting of a stable, steady flow
on the large scales and disorganized motions on the small scales.
This generic behavior is confirmed by numerical simulations of high
Reynolds' number flows in various settings.  For instance, a
freely-decaying flow with doubly-periodic boundary conditions relaxes
at long times to either a coherent dipole vortex or double shear layer
\cite{MMMO,SK}.  Similarly, a weakly driven and dissipated flow is
well approximated by a nearly steady coherent structure on the large
scales that changes slowly in response to the driving and dissipation
\cite{MH,GM}.

Quasi-geostrophic turbulence behaves in a similar fashion, producing
coherent structures on the large scales of motion within a potential
vorticity field that is turbulent on a range of small scales
\cite{Rhines}.  In a geophysical context such as the active
weather layer on Jupiter, robust mean flows of this kind are observed
in the form of persistent jets and spots \cite{Marcus}.  Numerous, but
less obvious, examples of long-lived mean flows with these general
characteristics are also found in the Earth's oceans and atmosphere
\cite{Pedlosky}.  Generically, these coherent structures are shear
flows or distributed vortices embedded in shear flows.

In this paper we study a statistical equilibrium theory of coherent
structures in two-dimensional or barotropic quasi-geostrophic
turbulence.  Several theories of this kind have been proposed and
their predictions have been analyzed in some detail; they include the
Onsager-Joyce-Montgomery theory of a point vortex gas
\cite{Onsager,JM,Kiessling,CLMP}, the Kraichnan energy-enstrophy
theory \cite{Kraichnan,SHH,Holloway,CF}, and the Miller-Robert theory
of a continuum vorticity field \cite{Miller,MWC,Robert,RS}.  A review
and critique of these various theories is given in \cite{Turkington}.
In that work it is shown that each of these theories relies upon some
explicit or implicit assumptions concerning the form of the random
vorticity field on the microscopic scale and that these different
assumptions lead to different predictions about the coherent structure
on the macroscopic scale.  These differences stem from the way in
which the generalized enstrophy invariants (the spatial higher moments
of vorticity) are included in the various theoretical models.  Unlike
the global invariants associated with the conservation of energy and
circulation, which are ``rugged'' invariants that depend on the large
scales of motion, the generalized enstrophy invariants are ``fragile''
in the sense that they are sensitive to the vorticity fluctuations on
the small scales.  

In the present paper we therefore consider a model in which the
fragile invariants are replaced by a given probability distribution on
the small-scale vorticity fluctuations, which we call the prior
distribution.  With respect to this underlying probabilistic
description of the vorticity field, we then impose the rugged global
invariants on the statistical equilibrium measure that defines the
model.  In this fashion we obtain a theory in which a single prior
distribution captures the microscopic effects and a few global
invariants control the macroscopic features.

Besides being more faithful to the continuum dynamics than the known
theories, this model is more easily adapted to realistic physical
situations.  On the one hand, there can be practical advantages to
having a model that utilizes only a few robust invariants, as has been
demonstrated in \cite{MH,DMT}.  On the other hand, a suitable prior
distribution can be fit directly to the one-point vorticity statistics
available from numerical simulations or physical data. Alternatively,
it can be inferred indirectly by comparing the predicted
vorticity-streamfunction profile with an observed profile.

In the context of a model of this kind, we have the choice of building
the equilibrium statistical measure from a canonical ensemble or from a
microcanonical ensemble with respect to the rugged invariants.  In
most applications of statistical mechanics these alternative
formalisms define equivalent theories that have identical equilibrium
states in the thermodynamic limit \cite{Balescu,Balian}. It is rather
surprising, therefore, to discover that in our models of coherent
structures the two ensembles are not always equivalent.  In fact, we
find that there are regimes in which the equilibrium states for the
microcanonical ensemble are entirely omitted by the canonical ensemble.
Moreover, numerical computations based on the microcanoncial model
show that these regimes often contain mean flows of great physical
interest.  In essence, the reason for this novel behavior lies in
the character of the statistical equilibrium models: they are local
mean-field theories in which the continuum limit is nonextensive, the
interactions are long-range, and the inverse temperature is negative.

Given that some equilibrium states for microcanonical model are not
realized by the corresponding canonical model, we are led to ask
whether these most probable states correspond to stable flows.  We
answer this question in the affirmative by proving that all
nondegenerate canonical and microcanonical equilibrium states define
nonlinearly stable, steady mean flows.  In the canonical model, these
results reduce to the well-known the Arnold stability theorems, which
rely on Lyapunov functionals constructed from the rugged invariants
and the information (negative entropy) functional associated with the
prior distribution \cite{Arnold,MP}.  In the microcanonical model,
however, these standard Lyapunov functionals are not positive definite
at those equilibrium states which are not realized by the canonical
model.  In the nonequivalent case we instead use a new class of
Lyapunov functionals to demonstrate the stability of the most probable
flows for the microcanonical model.  In this construction we introduce
a penalization of the standard functional with respect to the
microcanonical constraints that makes the resulting Lyapunov
functional positive definite at the microcanonical equilibrium states.
Such penalized functionals are identical with the so-called augmented
Lagrangians used in methods for constrained optimization
\cite{Bertsekas, Minoux}.

These results support our contention that the natural formulation of a
statistical equilibrium model of coherent structures is the one in
which conservation of generalized enstrophy is relegated to a prior
distribution, and conservation of energy and circulation are imposed
microcanonically.  From a mathematical standpoint, this model is
preferrable to the corresponding canonical model because, in general,
its family of equilibrium states is richer.  From a physical point of
view, the microcanonical conditions are pertinent because the energy
and circulation are trapped in the largest scales of motion, and hence
these rugged invariants are isolated from interactions with larger
systems or ignored degrees of freedom.  Reciprocally, the use of a
prior distribution on the vorticity, which amounts to a canonical
treatment of the generalized enstrophy invariants, acknowledges that
the statistical properties of the vorticity on the small scales are
determined by contact with a bath of unresolved turbulent
motions. Finally, our refined stability theorems ensure that the most
probable flows defined by the model are nonlinearly stable for any
admissible values of the microcanonical constraints, even when the
classical sufficient conditions for stability are not satisfied.

The paper is organized as follows.  In Section 2 we formulate a
general equilibrium statistical model that includes two-dimensional
and barotropic geostrophic turbulence with topography.  After
explaining the role of the prior distribution in the probabilistic
lattice model, we construct the canonical and microcanonical models,
respectively.  In Section 3 we then present the variational principles
for these two models in the continuum limit as the lattice spacing
tends to zero.  Our analysis makes use of large deviation techniques,
which are uniquely suited to derivations of this kind
\cite{Ellisbook,Ellisoverview}.  In particular, we introduce a
coarse-graining of the microscopic vorticity field and present the
fundamental large deviation principle that this process satisfies.  In
another paper we state and prove a general theorem that contains this
result as a special case \cite{EHT3}.  On the basis of this result, we
develop the variational principles governing the equilibrium
macrostates in the canonical and the microcanonical continuum models.
We give the complete proofs of the large deviation estimates needed to
justify these variational principles in a companion paper \cite{EHT1},
where we treat an general class of models defined in terms of local
mean-field interactions.  In Section 4 we turn to the equivalence of
ensembles questions, invoking ideas from convex analysis and
constrained optimization theory to obtain sharp and complete results.
A more general treatment of these issues is also presented in the
companion paper \cite{EHT1}.  In Section 5 we present the nonlinear
stability theorems, first reviewing the known theorems that pertain to
the canonical model and then developing the refinement of those
theorems that applies to the microcanonical model.  Finally, in Section
6 we display the results of some numerical solutions to the
microcanonical variational principle for barotropic shear flows over a
zonal topography.  In this physically interesting problem the
nonequivalence-of-ensembles behavior is quite conspicuous.

Our presentation throughout this paper is a synthesis of physical
modeling and mathematical analysis, which is intended to focus on the
conceptual aspects of the models we study.  With this goal in mind, we
omit many of the technical details and proofs, referring the reader to
our other papers \cite{BET,EHT1,EHT3} for those aspects.

%==========================================================================
\section{Formulation of the models}
%subsections:  
%2.1 Two-dimensional and geostrophic turbulence
%2.2 Generalized enstrophies and small-scale fluctuations
%2.3 Global invariants and large-scale motions

\noindent
{\bf 2.1 Two-dimensional and geostrophic turbulence.}  For the
microscopic dynamics that underlies our statistical equilibrium models
we adopt an equation of motion that contains as special cases the
governing equations for both purely two-dimensional turbulence and
barotropic quasi-geostrophic turbulence.  Namely, we consider the
nonlinear advection equation
\begin{equation} \label{qg}
\frac{\d Q}{\d t} \;+\; \frac{\d Q}{\d x_1}\frac{\d \psi}{\d x_2} \;-\;
 \frac{\d Q}{\d x_2}\frac{\d \psi}{\d x_1} \;=\; 0 \, ,
\end{equation}
in which $Q=Q(x_1,x_2,t)$ and $\psi=\psi(x_1,x_2,t)$ are real scalar fields
related by the elliptic equation
\begin{equation} \label{pv}
Q \;=\; - \Delta \psi + r^{-2} \psi + b \, .
\end{equation}
In this defining equation, $\Delta = \d ^2/\d x_1^2 + \d ^2/\d x_2^2 $
denotes the Laplacian on $\R^2$; $r$ is a given positive constant
which may be infinity; and $b=b(x_1,x_2)$ is a specified
continuous function.  The flow velocity field $v$ is nondivergent and
is determined from the streamfunction $\psi$ by $v=(\d\psi/\d
x_2,-\d\psi/\d x_1)$.  For the sake of definiteness, we take the flow
domain to be a channel
\begin{equation} \label{domain}
\X \;=\; \{ x=(x_1,x_2) \, : \, |x_1| < \ell_1/2 \, , \, |x_2| < \ell_2/2 \, \}
\end{equation}
with a period length $\ell_1$ and finite width $\ell_2$.  The boundary
conditions for ideal flow in such a channel are achieved by setting
$\psi = 0$ on the walls $x_2= \pm \ell_2/2$ and by imposing
$\ell_1$-periodicity in $x_1$.

Equations (\ref{qg})-(\ref{pv}) reduce to the Euler equations
governing incompressible, inviscid flows in two dimensions when $r=
\infty$ and $b=0$. In this case $Q$ coincides with the vorticity
$\omega = \d v_2 / \d x_1 - \d v_1 / \d x_2$.  In such a flow the
conservation of momentum is equivalent to the exact rearangement of
vorticity $\omega$ under the area-preserving flow maps for the
velocity field $v$ induced instantaneously by $\omega$.

When a finite $r$ and a nonvanishing $b$ are included in
(\ref{qg})-(\ref{pv}), these general equations contain the standard
equations governing a shallow rotating layer of homogeneous
incompressible inviscid fluid in the limit of small Rossby number.  In
the geophysical literature where these equations are derived and
discussed \cite{Pedlosky}, the nondimensionalized spatial variables
$(x_1,x_2)$ are written as $(x,y)$ and the geostrophic streamfunction
$\psi$ is replaced by $-\psi$, which also represents the
nondimensionalized free-surface perturbation.  Under appropriate
quasi-geostrophic scalings and up to first-order in the Rossby number,
the flow is nondivergent and its potential vorticity $Q$, defined by
(\ref{pv}), is advected by the flow according to (\ref{qg}).  The
inhomogeneous term in (\ref{pv}) is given by $b = \beta y + h$, where
$\beta$ is the gradient of the Coriolis paramter $f=f(y)$ and $h$ is
the height of the bottom topography.  The constant $r$ in (\ref{pv})
is the Rossby deformation radius $r= \sqrt{g H_0} / f_0$, which is
determined by the gravitational acceleration $g$, the mean fluid depth
$H_0$, and a mean value $f_0$.  We refer the reader to the literature
for a complete discussion of these fundamental equations and their
properties \cite{Pedlosky}.

The general equations (\ref{qg})-(\ref{pv}) also contain the governing
equations for the so-called 1-1/2 layer model, in which a shallow
upper layer lies on a deep lower layer of denser fluid whose motion is
unaffected by that in the upper layer.  Besides having oceanographic
applications, this model is often used to describe the observed
weather layer of the Jovian atmosphere \cite{IC,Marcus}.  In the
applications to Jupiter, the lower layer flow is assumed to be steady,
zonal and geostrophically balanced.  Then the potential vorticity for
the active upper layer is given by (\ref{pv}) with $b = \beta y -
r^{-2} \psi_2(y)$, where $\psi_2$ denotes the streamfunction for the
flow in the lower layer.  In this way, the deep flow produces an
effective bottom topography.  The appropriate Rossby scale $r$ is
determined as in the single layer model, except that a reduced gravity
$g'$ is used.  With these choices, (\ref{qg})-(\ref{pv}) govern
the quasi-geostrophic dynamics of the shallow upper layer.

>From the point of view of statistical equilibrium theory, the
underlying continuum dynamics dictated by (\ref{qg})-(\ref{pv}) serves
as a mechanism for mixing the scalar field $Q$ subject to the
constraints imposed by the various conserved quantities for that
dynamics.  Indeed, the equilibrium statistical models that we study
are constructed by postulating that the underlying dynamics is ergodic
with respect to the ideal invariants.  This ergodic hypothesis is not
expected to be universally valid.  Nevertheless, numerous observations
and simulations of two-dimensional and geostrophic turbulence show
that typically the self-induced straining of the advected field $Q$
leads to an effective randomization of $Q$.  For instance, in a free
evolution from a generic smooth field $Q^0$, $Q$ develops local
finite-amplitude fluctuations on a range of small scales as time
progresses.  This behavior is related to the direct cascade of
enstrophy to small scales.  At the same time, $Q$ tends to organize
into coherent vortices at the large scales, and these vortices
gradually merge into a final steady state.  This dual behavior is
associated with the inverse cascade of energy to large scales.  The
goal of the statistical equilibrium models is to characterize the
typical steady mean flows that persist on the large scales without
resolving the small scales of motion.  The validity of these models
must be checked a posteriori from their predictions, since a priori
tests or proofs of the ergodic hypothesis are generally not feasible.

The conserved quantities associated with (\ref{qg})-(\ref{pv}) are the
total energy $H$, the total circulation $C$, and a family of generalized
enstrophies $A$, given by,
\begin{equation} \label{energy_psi}
H \;=\; \frac{1}{2} \int_{\X} 
\left[ \left(\frac{ \d \psi}{ \d x_1}\right)^2  \,+\,
    \left(\frac{ \d \psi}{ \d x_2}\right)^2  \;+\;
       r^{-2} \psi^2 \right] \, dx \, ,
\end{equation}
\begin{equation} \label{circulation}
C \;=\;  \int_{\X} [ \, Q  - b \, ] \, dx \, ,
\end{equation}
\begin{equation} \label{enstrophy}
A \;=\; \int_{\X} a(Q)  \, dx \, ,
\end{equation}
where $a$ is an arbitrary, sufficiently smooth, real function on the
range of $Q$.  In addition, the $x_1$-component of linear impulse
(momentum) $M$ is also conserved in the channel geometry that we
consider; it is given by
\begin{equation} \label{impulse}
M \;=\; \int_{\X} x_2 [ Q  - b ] \, dx \, .
\end{equation}
We note that the expression $Q - b =\zeta + r^{-2} \psi$ appearing
in the circulation and impulse integrals is a sum of the relative vorticity,
$\zeta = - \Delta \psi$, and the vortex stretching term, $r^{-2} \psi$,
due to deformation of the free-surface.  

While each of these quantities is precisely conserved by the continuum
dynamics, the role that $H$, $C$ and $M$ play in the statistical
equilibrium models differs dramatically from that played by the
nonlinear enstrophies $A$.  This crucial difference is a consequence
of the fact that the generalized enstrophy invariants are sensitive to
the small-scale structure of $Q$, while the energy, circulation, and
impulse invariants depend on the large scales of motion.  For this
reason, in the following two subsections we formulate the various
models by first defining the probabilistic structure of the small
scales and then introducing the conditioning determined by the global
invariants for the large scales.

\vspace{.2in}
\noindent
{\bf 2.2 Generalized enstrophies and small-scale fluctuations.} In
order to define our continuum models, we first replace the infinite
dimensional phase space of continuum vorticity fields $Q$ by a
sequence of finite dimensional phase spaces and then take an
appropriate continuum limit.  To this end, we introduce a lattice $\L$
on the domain $\X$ having $n$ sites and construct a probabilistic
lattice model for each $n$.  It suffices to use a uniform intersite
spacing in both the $x_1$ and $x_2$ directions; say, a dyadic
partition of the intervals $-\ell_1/2 < x_1 < \ell_1/2$ and $-\ell_2/2
< x_2 < \ell_2/2$ into $2^{m_1}$ and $2^{m_2}$ equal parts, so that
$n=2^{m_1+m_2}$.  The domain $\X$ then consists of the disjoint union
of $n$ microcells $M(s)$ indexed by the sites $s$ in the lattice $\L$.
The phase space for the lattice model is the product space
$\Omega_n=\R^n$, the microstates in the lattice model being points in
$\Omega_n$.  We identify these micrstates with vorticity fields $Q$
that are piecewise-constant relative to $\L$; that is, $Q(x)=Q(s)$ for
all $x \in M(s), \; s \in \L $.  For the sake of simplicity, we shall
use the same notation for the continuum field $Q(x), \; x \in \X,\, $
governed by the underlying partial differential equations and the
discretized field $Q(s), \; s \in \L, \,$ in the lattice model.

The small-scale fluctuations of the microstates in the lattice model
are described by the product measure
\begin{equation}  \label{pi_n}
\Pi_n(dQ) \;=\; \prod_{s \in \L} \rho (dQ(s))  \;\;\;\;\;\;
\mbox{ on } \;\; \Omega_n \, , 
\end{equation}
in which $\rho (dy)$ is a given probability distribution on $\R$.
Here and throughout the paper, $y$ denotes a real variable running
over the range of $Q$.  With respect to the probability distribution
$\Pi_n$ the microscopic fields $Q$ consist of $n$ independent,
identically distributed random variables over the $n$ microcells in
the lattice.  We refer to the common distribution $\rho$ as the {\em
prior distribution}, signifying that it describes the statistical
properties of the microstate $Q$ before the conditioning due to the
rugged invariants is imposed.

When $\rho(dy)=e^{-a(y)}dy, \; y \in \R$, for some continuous function
$a$ on $\R$, the product measure $\Pi_n$ in (\ref{pi_n}) coincides
with canonical Gibbs measure with respect to $A \doteq \frac{1}{n}
\sum a(Q(s))$, which is the discetization of the generalized
enstrophy integral $A=\int a(Q)dx$.  That is,
\[ 
\Pi_n(dQ) \;=\; e^{ - n A_n(Q)} \, \prod_{s \in \L} dQ(s) 
          \;=\; \prod_{s \in \L} e^{ - a(Q(s))} \, dQ(s) \, .
\] 
In light of this identity, the role of $\Pi_n(dQ)$ in the lattice
model is evident from the general principles of statistical mechanics:
it is the most probable distribution on $\Omega_n$ with respect to the
phase volume $dQ=\prod dQ(s)$ that is consistent with the conservation
of generalized enstrophy $A_n$.  Typically, this characterization of the
canonical ensemble is justified by two dynamical properties: 1) the
invariance under the phase flow of the phase volume $dQ$; 2) the
dynamical invariance of the function $A_n$.  In the models we study,
however, a lattice dynamics discretizing the underlying continuum
dynamics for which these two properties hold is not known.
Consequently, it is necessary to treat the construction of the product
measure $\Pi_n(dQ)$ as a modeling issue, justifying its choice on
whatever theoretical results are available and whatever practical
considerations are at hand.   

The principal reason for preferring the canonical ensemble $\Pi_n(dQ)$
to the corresponding microcanonical ensemble is the sensitivity of the
generalized enstrophies $A$ to small-scale motions.  In physical terms
$\Pi_n(dQ)$ describes a random field $Q$ on the lattice $\L$ in which
there is a coupling between the fluid motions on scales resolved by
the lattice and the unresolved turbulence on smaller scales.  As in
standard statistical equilibrium theory, the canonical formulation is
appropriate to a system coupled to a reservoir, or thermal bath
\cite{Balescu,Balian}.  The prior distribution $\rho$ that
parametrizes $\Pi_n(dQ)$ is effectively a generalized inverse
temperature for the potential vorticity fluctuations on the lattice
microscale.  By contrast, the microcanonical ensemble based on a
(finite or infinite) family of generalized enstrophies $A_n$ enforces
the exact conservation of each $A_n$ on the lattice, inhibiting the
exchange of generalized enstrophy between the resolved scales and the
unresolved scales.  The well-known flux of enstrophy to small scales
therefore invalidates the microcanonical formulation.

Statistical equilibrium theories of the long-time average behavior of
solutions to (\ref{qg})-(\ref{pv}) have tended to emphasize the
microcanonical formulation.  Originally, Miller \cite{Miller,MWC} and
Robert\cite{Robert,RS} independently constructed a model by assuming
that the exact rearrangement of vorticity under the continuum dynamics
is imitated on the lattice $\L$ by an unspecified lattice
dynamics. Under this assumption all generalized enstrophies $A_n$ are
exactly conserved in the lattice model.  This approach produces a
well-defined model in which the complete family of vorticity
invariants is imposed microcanonically.  Later, Turkington
\cite{Turkington,BET} criticized the assumption made in the
Miller-Robert model and formulated a modification of it that is
derived instead from the underlying exact continuum dynamics on $\X$.
In the Turkington model, the evolution of the continuum vorticity
field is observed on the lattice $\L$ by averaging over the scales
smaller than the lattice microscale, and consequently the family of
equality constraints on all generalized enstrophies imposed in the
Miller-Robert model is replaced by a weaker family of inequality
constraints on all convex enstrophies.  This approach results in a
model that accounts for the partial loss of the nonlinear enstrophies
to submicroscale fluctuations.  Among statistical equilibrium models
that associate a final coherent state with a given initial state this
model is the most faithful to the underlying ideal continuum dynamics.  

For the reasons mentioned above, however, a canonical formulation with
respect to the generalized enstrophies usually furnishes a more
appropriate physical model than a microcanonical formulation.
Moreover, the equilibrium equations for the Turkington model are
isomorphic to those for the canonical model with a prior distribution
(\ref{pi_n}) under the identification $\rho(dy) = e^{ -a(y)} dy$; in
the microcanonical case the function $a$ is determined by the
Kuhn-Tucker multipliers for the family of convex enstrophy
inequalities, while in the canonical case it is prescribed
\cite{Turkington,BET}.  Whether all microcanonical equilibria are
realized as canonical equilibria is not known.

In practical applications these statistical equilibrium models are
used to produce families of most probable large-scale flows that
coexist with other complex mechanisms influencing the small-scale
motions.  Under these circumstances the canonical ensemble
(\ref{pi_n}) is often desirable because the prior distribution $\rho$
can be used to model the one-point probability distribution of the
vorticity fluctuations.  On the other hand, the constraints on
generalized enstrophies, or potential vorticity moments, are of
dubious relevance in these realistic situations.  For instance, in
two-dimensional turbulence with weak driving and small dissipation it
is possible to invoke a statistical equilibrium model as an adiabatic
approximation to the evolution of the large-scale structure
\cite{MH,GM,DM2}.  In these applications only the lowest-order moments
of vorticity are sufficiently robust to be retained in the model.
Similarly, comparisons with direct numerical simulations of
freely-decaying turbulence show good agreement with the predictions of
the model only when the higher-order moments of vorticity is altered
to account for dissipation \cite{BSP}. These tests show that it is
necessary to take a prior distribution that is compatible with the
relaxed final state.  In the context of geostrophic turbulence, the
modeling of the turbulent small scales is further complicated by the
possible effects of nonvanishing Rossby and Froude numbers
\cite{PMSF}. Given the asymptotic nature of the quasi-geostrophic
equations themselves, it is reasonable to fit the prior distribution
to available data.  In Section 6, we briefly indicate how this
empirical approach can be used to formulate a model of zonal jets in a
Jovian atmosphere.

For the purposes of our general discussion throughout Sections 3, 4
and 5, we let the prior distribution $\rho$ be an arbitrary
probability distribution on $\R$ subject only to the decay condition
(\ref{prior_growth}), and we base all of our models on the canonical
ensemble (\ref{pi_n}) parametrized by such $\rho$.  This simple choice
of the product measure $\Pi_n(dQ)$ is natural in the context of
statistical equilibrium theory.  Any better choice would require a new
theory of the correlation structure of turbulent scales, derived
presumably from nonequilibrium considerations.

\vspace{.2in}
\noindent
{\bf 2.3 Global invariants and large-scale motions.} The statistical
equilibrium lattice models that we consider are constructed by
imposing the global invariants $H$ and $C$ on the product measure
$\Pi_n(dQ)$.  In this construction we can consider either the
canonical or the microcanonical ensemble with respect to these
invariants.  A main goal of this paper is to investigate the
equivalence or nonequivalence of these two different ensembles.
Accordingly, we now proceed to formulate these canonical and
microcanonical models.

The canonical model is defined by the Gibbs distribution 
\begin{equation} \label{can_distr}
P_{n,\beta,\gamma} (dQ) \;=\; Z_n(\beta,\gamma)^{-1} \, 
                       \exp ( - n \beta H_n(Q) - n \gamma C_n(Q) ) \, 
                            \Pi_n(dQ) \, ,
\end{equation}
and is parametrized by $\beta, \gamma \in R$, which play the roles of
``inverse temperature'' and ``chemical potential,'' respectively.  The
partition function
\[
Z_n(\beta,\gamma) \;=\; 
   \int_{\Omega_n} \exp ( - n\beta H_n(Q) - n\gamma C_n(Q) ) \, \Pi_n(dQ) \, 
\]
normalizes the probability distribution $P_{n,\beta,\gamma} (dQ)$ on
$\Omega_n$.  We use the traditional notation $\beta$ for inverse
temperature even though this symbol overlaps with that used in the
geophysical literature for the gradient of the Coriolis parameter; we
expect that the distinction will be clear enough from context.

The microcanonical model is defined by the conditional distribution
\begin{equation} \label{micro_distr}
P_n^{E,\Gamma} (dQ) \;=\; \Pi_n \left\{ \, dQ \; | 
                        \; H_n(Q) = E, \; C_n(Q) = \Gamma \, \right\} \, ,
\end{equation}
at given values $E$ and $\Gamma$ of the global invariants.  For
technical reasons, it is necessary to replace the exact equality
$H_n=E$ in (\ref{micro_distr}) by a containment $H_n \in [
E-\epsilon,E+\epsilon ]$ for a small finite $\epsilon > 0$ and
similarly for the exact equality $C_n=\Gamma$.  For the sake of
clarity of exposition, however, we will ignore this technical point
and set $\epsilon=0$ throughout our discussion, leaving the obvious
adjustments to the reader.

The functionals $H_n$ and $C_n$ in ({\ref{can_distr}) and
({\ref{micro_distr}) are the lattice versions of the functionals $H$
and $C$ defined on the continuum field $Q$ in (\ref{energy_psi}) and
(\ref{circulation}), respectively. $H_n$ and $C_n$ act on $\Omega_n$ by
identifying each microstate $Q \in \Omega_n$ with the corresponding
piecewise-constant function $Q \in L^{2}(\X)$, and by evaluating the
functionals $H$ and $C$ on that field; the corresponding solution
$\psi$ to (\ref{pv}) then determines $H(Q)$.  Some straightforward
calculations show that they have the explicit expressions
\[ 
H_n(Q) \;=\; \frac{\ell_1^2 \ell_2^2}{2n^2} 
     \sum_{s \in \L} \sum_{s' \in \L} g_n(s, s') Q(s) Q(s') \;-\;
\frac{\ell_1 \ell_2}{2n} \sum_{s \in \L} h_n(s) Q(s) \, ,
\]
\[
C_n(Q)  \;=\; \frac{\ell_1 \ell_2}{n} \sum_{s \in \L} Q(s) - b(s) \, ,
\]
where $ g_n(s, s')$ is the average over $M(s) \times M(s')$ of the
Green function $g(x,x')$ defined by $(-\Delta + r^{-2}) g =
\delta(x-x')$, and $h_n(s)$ is the average over $M(s)$ of the solution
$h(x)$ to $(-\Delta + r^{-2} ) h = b(x)$; both $g(x,x')$ and $h(x)$
satisfy the boundary conditions on $\d \X$ imposed on $\psi(x)$.  The
lattice energy $H_n$ consists of a quadratic self-interaction term
with a potential $g_n$ and a linear term involving $h_n$ that represents
interaction with the bottom topography.

It is important to note that the vortex self-interactions governed by
$H_n$ are long-range, being determined essentially by the Green
function $g(x,x')$ for the partial differential operator $-\Delta +
r^{-2} $ on $\X$.  This property, combined with the form of the
product prior distribution $\Pi_n(dQ)$, gives these statistical
equilibrium models their character as local mean-field theories.
Moreover, the long-range interactions imply that the energy function
$H_n$ is a rugged invariant, meaning that it is not sensitive to the
small-scale structure of the vorticity field.  Indeed, $H_n$ depends
only on the local average of $Q$ in a neighborhood of any point, and
therefore it is well approximated a spatial coarse-graining of $Q$.
The same properties are shared by $C_n$ because it is a linear
function of $Q$.  By contrast, as stressed in the preceding
subsection, all nonlinear enstrophies $A_n$ are fragile invariants in
the sense that they cannot be approximated by their values on a
coarse-grained state.

The canonical parameters $\beta$ and $\gamma$ are scaled by a factor
$n$ in (\ref{can_distr}).  This scaling ensures that, in the continuum
limit as $n \rightarrow \infty$, the mean values $ \langle H_n
\rangle$ and $ \langle C_n \rangle$ with respect to this canonical
ensemble tend to finite limits, and that the variances of $H_n$ and
$C_n$ around these mean values tend to zero.  The canonical ensemble
(\ref{can_distr}) thus produces equilibrium states having finite total
energy and total circulation in the continuum limit, and hence it is
compatible with the microcanonical ensemble (\ref{micro_distr}) in
which $E$ and $\Gamma$ are fixed and finite as $n \rightarrow \infty$.
We note that, while this scaling of the parameters determining the
canonical ensemble is natural in these local mean-field models, it
results in a nonextensive continuum limit that is different from the
usual thermodynamic limit \cite{MWC}.

The linear impulse invariant $M$, which is associated with the
translational symmetry of the channel domain, can also be included in
either the canonical or the microcanonical ensembles.  For the sake of
clarity, however, we ignore it in our development.  If $M$ is treated
canonically, then the energy function $H_n$ is simply replaced by $H_n
+ U M_n$, where $(U,0)$ is the velocity of a given uniform zonal flow.
Alternatively, if $M$ is imposed microcanonically, then $U$ is
determined implicitly.  In either formulation, the analysis of the
impulse constraint is the same as that of the circulation constraint,
which is also linear in $Q$.

%==========================================================================
\section{Maximum entropy principles}
%subsections: 
%3.1 Coarse-grained process
%3.2 Canonical model
%3.3 Microcanonical model
%

\noindent
In this section we investigate the continuum limit of the canonical
and microcanonical models constructed in the preceding section, and we
thereby derive the maximum entropy principles which characterize the
most probable states for those models.  Our analysis of the continuum
limit relies on the powerful methods of the theory of large deviations
\cite {Ellisbook,DZ}.  First, we establish a large deviation principle
for a certain coarse-graining of the potential vorticity field $Q$
with respect to the product prior distribution $\Pi_n(dQ)$.  With this
basic result in hand, we then analyze the canonical ensemble
$P_{n,\beta,\gamma}$ and the microcanonical ensemble $P_n^{E,\Gamma}$,
and establish large deviation principles for the coarse-grained field
with respect to each of these ensembles.  In this way we obtain a
variational characterization of the equilibrium macrostates for each
model.

\vspace{.2in}
\noindent
{\bf 3.1 Coarse-grained process.}  We now introduce a macroscopic
description of the potential vorticity field that complements the
microscopic description inherent in the lattice model.  We take
the space of macrostates $q$ to be the Hilbert space $L^2(\X)$ with
the usual norm $\|q\|^2 = \int_{\X} q^2 dx$.  This natural and
convenient choice requires us to impose a certain decay condition on
the prior distribution $\rho$.  Specifically, we assume that there
exists $\delta >0$ such that
\begin{equation} \label{prior_growth}
\int_R \exp \left( \frac{\delta}{2} |y|^2 \right) \, \rho(dy)
\; < \, \infty \, .   
\end{equation}
Since this decay condition holds for most prior distributions of
interest, including compactly supported and Gaussian distributions, we
adopt it for the sake of simplicity throughout Sections 3, 4 and 5.
In Section 6, however, we relax it for a particular prior distribution
used in the numerical example.

In order to establish the connection between the microscopic and
macroscopic levels of description, we define a certain {\em
coarse-grained process} as follows.  Partition the domain $\X$ into
$\ntilde = 2^{r_1+r_2}$ macrocells $\X_{j_1,j_2}$, with $r_1 \ll m_1$,
$r_2 \ll m_2$, and $j_1=1, \ldots , 2^{r_1}$, $j_2=1, \ldots ,
2^{r_2}$.  This partition represents a coarsening of the lattice $\L$
that defines the phase space $\Omega_n$; each of the $\ntilde$
macrocells $\X_{j_1,j_2}$ contains $n/\ntilde$ sites of $\L$.  Now,
let $\Qtilde$ be the $L^2(\X)$-valued stochastic process defined by
averaging the random microstate $Q$ over each macrocell; namely,
\begin{equation} \label{coarse} 
\Qtilde (x) \;=\; \frac{\ntilde}{n} \sum_{s \in \X_{j_1,j_2} } Q(s) 
\;\;\;\;\;\;\;\; \mbox{ for all } \;\; x \in \X_{j_1,j_2} \, ,
\end{equation}
Clearly, $\Qtilde$ is piecewise constant with respect to the partition
of $\X$ into macrocells.  The coarse-grained process $\Qtilde$ takes
values in the space of macrostates $L^2(\X)$.   

In what follows we shall be interested in a double limit in which both
$n \rightarrow \infty$ and $\ntilde \rightarrow \infty$, with
$n/\ntilde \rightarrow \infty$.  We refer to this double limit as the
{\em continuum limit}.  In order to deduce the limiting behavior of
$\Qtilde$ under either the canonical ensemble (\ref{can_distr}) or the
microcanonical ensemble (\ref{micro_distr}), we first estabilish a
basic theorem that describes its behavior with respect to $\Pi_n(dQ)$.
The formulation of this theorem requires some definitions, which
we now state.     

Associated with the prior distribution $\rho$ is its {\em cumulant
 generating function}
\begin{equation} \label{cgf}
f(\eta) \;=\; \log \int_{R} \exp ( \eta y ) \, \rho(dy)
\;\;\;\;\;\;\;\;\;\; ( \, \eta \in R \, ) \, .
\end{equation}
In view of the decay condition (\ref{prior_growth}), $f(\eta)$ is
defined and continuous for all $\eta \in R$.  Moreover, $f$ is
convex function.  The convex function $i$ conjugate to $f$, namely,
the Legendre-Fenchel transform of $f$, is defined by
\begin{equation} \label{i}
i(y) \; = \; \sup_{\eta} \, [ \, \eta y - f( \eta) \, ] \;\;\;\;\;\;\;\;\;\;
( \, y \in R \, ) \, .
\end{equation}
It is known that $i$ achieves its unique minimum value of $0$ over
$\R$ at $\bar{y} \doteq \int y \rho(dy)$.  The reader is referred to
\cite{DZ,Ellisbook} for these definitions and properties.

In terms of these standard constructions, we define the {\em
information functional}
\begin{equation}   \label{info}
I(q) \;=\; \int_{\X} i(q(x)) \, dx  
\;\;\;\;\;\;\;\;\;\;\;  ( \, q \in L^2(\X) \, ) \, .
\end{equation}
In the terminology of large deviation theory, $I$ is a convex rate
function; that is, it is a convex, lower semi-continuous functional
mapping $L^2(\X)$ into the extended interval $[0,+\infty]$.  In fact,
the information functional $I$ is the rate function for the basic
large deviation principle satisfied by the coarse-grained process
$\Qtilde$ with respect to the product prior distribution $\Pi_n(dQ)$.
In the following theorem we state a simplified version of this large
deviation principle.  In another paper \cite{EHT3}, we state and prove
the general version.

\vspace{.2in}
\noindent
{\bf Theorem 1.} For any Borel subset $B$ of $L^2(\X)$ that is a
continuity set for the rate function $I$, the following double limit
holds:
\begin{equation} \label{base_ldp}
\lim_{\ntilde \rightarrow \infty} \lim_{n \rightarrow \infty} 
\frac{1}{n} \log \Pi_n \{ \Qtilde \in B \} \;=\; - I(B)  \, , 
\end{equation}
where $I(B) \doteq \inf \{ I(q) : \, q \in B \}$.   
\vspace{.2in}

Here, we use the notion of a {\em continuity set} $B$ for $I$ to
assert simply a double limit rather than the standard pair of large
deviation upper and lower bounds for closed sets $B$ and open sets
$B$, respectively.  By a continuity set for the information functional
(\ref{info}) we mean any Borel set $B \subset L^2(\X)$ with the
property that $I(B^0)=I(\bar{B})$, where $B^0$ is the interior of $B$
and $\bar{B}$ is the closure of $B$.  Under suitable conditions on
$\rho$, such as (\ref{prior_growth}), the continuity sets of $I$ are
rich enough to encompass the sets that arise in practical applications
of the result.  The double limit (\ref{base_ldp}) then conveys
conceptually the content of the rigorous large deviation principle
given in \cite{EHT3}.  The proof relies essentially on the classical
Cram\'{e}r theorem for sample means of independent and identically
distributed random variables \cite{DZ,Ellisbook}.  Roughly speaking,
the theorem follows by applying Cram\'{e}r's theorem to the local
average that defines the coarse-grained process $\Qtilde$ over each
macrocell $\X_{j_1,j_2}$, and then by integrating the results for each
macrocell over the entire domain $\X$.

The asymptotic formula (\ref{base_ldp}) give an exponential-order
corrections to the law of large numbers behavior of the coarse-grained
process $\Qtilde$.  That is, finite departures of $\Qtilde$ from its
mean value, the constant $\bar{y} \doteq \int y \rho(dy)$, have
exponentially small probability as $n \rightarrow \infty$.  If we take
$B = \{ q \in L^2(\X) \, : \, \| q - \bar{y} \| \ge \delta > 0 \} $ in
(\ref{base_ldp}), then we have, for large $n$, $\ntilde$ and
$n/\ntilde$,
\[
\Pi_n \{ \Qtilde \in B \} \; \le \; e^{ - n I(B)/2 } \, ;
\]
in the formula $I(B) > 0$ for any finite $\delta$, while $I(\bar{y}) =0$.

We may summarize the content of Theorem 1 in the formal asymptotic
statement that, for any macrostate $q \in L^2(\X)$,
\begin{equation} \label{base_asymptotic} 
\Pi_n \{ \Qtilde \approx q \} \; \sim \; e^{ -n I(q) } \;\;\;\;\;\;\;\;\;\;
\mbox{ in the continuum limit. } 
\end{equation}
Here, the symbol $\approx$ means close in the strong topology of
$L^2(\X)$.  The equivalence between this formal statement and the
precise result (\ref{base_ldp}) can be seen by using balls $B_r(q)$ of
arbitrarily small radius $r$ centered at $q$, and the fact that $I(q)
= \lim_{ r \rightarrow 0} I(B_r(q))$.  This asymptotic expression also
provides the heuristic interpretation of the rate functional $I$ as a
negative entropy.  Indeed, $-I(q)$ quantifies the multiplicity of the
microstates that correspond under the coarse-graining to a macrostate
$q$.  Equivalently, $I(q)$ represents the information lost in going from
the microscopic to the macroscopic level of description.

\vspace{.2in}
\noindent
{\bf 3.2 Canonical model.} We now turn to the analysis of the
statistical equilibrium model governed by the canonical ensemble
(\ref{can_distr}).  The following theorem characterizes the continuum
limit for that model, using the asymptotics for the coarse-grained
process $\Qtilde$.

\vspace{.2in}
\noindent
{\bf Theorem 2.}  With respect to the canonical ensemble
$P_{n,\beta,\gamma}(dQ)$, the coarse-grained process $\Qtilde$
satisfies the double limit
\begin{equation} \label{can_ldp}
\lim_{\ntilde \rightarrow \infty} \lim_{n \rightarrow \infty} 
\frac{1}{n} \log P_{n,\beta,\gamma} \{ \Qtilde \in B \} \;=\; 
                  - I_{\beta,\gamma}(B) \, , 
\end{equation}
for any Borel subset $B$ of $L^2(\X)$ that is a continuity set
for $I_{\beta,\gamma}$; in this formula, 
\begin{equation} \label{can_info}
 I_{\beta,\gamma}(q) \; \doteq \; I(q) + \beta H(q) + \gamma C(q) 
                                   - \Phi(\beta,\gamma) \, ,
\end{equation}
where 
\begin{eqnarray} \label{free_energy}
\Phi(\beta,\gamma)  & \doteq &  
\min_{ q \in L^2(\X)} \; [ \, I(q) + \beta H(q) + \gamma C(q) \, ]  \\
                    &  =  & 
 - \lim_{n \rightarrow \infty} 
         \frac{1}{n} \log Z_n(\beta,\gamma) \, .   \nonumber
\end{eqnarray}
\vspace{.15in}

The proof of this theorem is indicated in our companion paper
\cite{EHT1}.  The key idea is to represent the interaction functions
$H_n$ and $C_n$ in the Gibbs measure (\ref{can_distr}) in terms of the
coarse-grained process and the corresponding continuum functionals $H$
and $C$.  This representation is provided by the following
approximations
\begin{equation} \label{representation}
H_n(Q) \,=\, H(\Qtilde) \,+\, o(1) \, , \;\;\;\;\;\;
C_n(Q) \,=\, C(\Qtilde) \,+\, o(1) \, ,
\end{equation}
in which the $o(1)$ errors are uniformly small over $Q \in \Omega_n$.
Here, and henceforth, we evaluate the functionals $H$ and $C$ defined
in (\ref{energy_psi}) and (\ref{circulation}) on macrostates $q \in
L^2(\X)$.  The streamfunction $\psi$ corresponding to any such $q$ is
the solution to $-\Delta \psi + r^{-2} \psi + b = q$ in $\X$ with
appropriate boundary conditions on $\partial \X$; that is,
$\psi=G(q-b)$, where $G$ denotes the Green operator for $-\Delta +
r^{-2}$:
\begin{equation} \label{green_operator}
Gz (x) \;=\; \int_{\X} g(x,x') \, z(x') \, dx'   \;\;\;\;\;\;\;\;
(\, z \in L^2(\X) \,) \, .   
\end{equation}
The approximations (\ref{representation}) express the fundamental fact
that the global invariants $H$ and $C$ are not sensitive to the
small-scale fluctuations of the microstate $Q$, being almost unchanged
by the local averaging that defines the coarse-grained process.  The
quadratic self-interaction term in $H$ has this property because it is
defined by the long-range interaction function $g(x,x')$.  $C$ and the
term in $H$ arising from interaction with the topography have this
property because they are linear (affine).  With the representations
in hand, the large deviation limit (\ref{can_ldp}) and the limit in
(\ref{free_energy}) can be established by general methods, namely, the
Laplace method for the asymptotics of large deviation-type
expectations.  As the proofs are very similar to those already given
in \cite{BET}, we omit them here.

>From the point of view of predicting the coherent states in a
turbulent fluid, the essential content of the large deviation
principle for the canonical ensemble lies in the {\em canonical
information functional} $I_{\beta,\gamma}$.  According to
(\ref{can_ldp}), the most probable macrostates $q $ are those at
which $I_{\beta,\gamma}(q)$ achieves its minimum value of $0$.  For
this reason, we define the {\em set of equilibrium states} associated
with given canonical parameters $\beta,\gamma \in R$ to be
\begin{equation} \label{can_equil}
\E_{\beta,\gamma} \; \doteq \; 
                     \{ q \in L^2(\X) \, : \, I_{\beta,\gamma}(q) = 0 \, \}
                  \;=\; \arg \min \, [ \, I + \beta H + \gamma C  \, ] \, .
\end{equation}
Any macrostate $q$ that does not lie in $\E_{\beta,\gamma}$ has an
exponentially small probability of being observed as a coarse-grained
state in the continuum limit; indeed, for such a macrostate
$I_{\beta,\gamma}(q) \ge \delta $ for some positive $\delta $, and
therefore the large deviation principle implies that for large
$n$ and $\tilde{n}$ 
\[
P_{n,\beta,\gamma} \{ \Qtilde \approx q \} \; \sim \; 
                    e^{ - n I_{\beta,\gamma} (q) } 
                          \; \le \; e^{ - n \delta }      \, .
\]
In light of this sharp estimate, we see that the equilibrium
macrostates in $\E_{\beta,\gamma}$ are overwhelmingly most probable
among all possible coarse-grained states of the turbulent system.
Consequently, the main predictions of the statistical equilibrium
theory in its canonical form are derived by solving the unconstrained
minimization problem whose objective functional is $I + \beta H +
\gamma C $.  The existence of at least one equilibrium state $\qbar$
in $\E_{\beta,\gamma}$ for each given $\beta,\gamma \in R$ can be
deduced readily by the direct methods of the calculus of variations.
In general, $\E_{\beta,\gamma}$ may contain more than one macrostate,
in which case the statistical equilbrium model exhibits a phase
transition.

Let us now display the first-order conditions for the
variational problem whose solutions are the equilibrium states in
the canonical model.  At a given solution $\qbar \in \E_{\beta,\gamma}$,
there holds
\begin{eqnarray}  \label{first_order}
0 & = & \delta (\, I + \beta  H + \gamma  C \,)(\qbar)  \\
  & = & \int_{\X} \, [ \, i'(\qbar) + \beta \psibar + \gamma \, ] 
                                          \; \delta q \; dx \; , \nonumber
\end{eqnarray}
where $\psibar$ is the streamfunction corresponding to  $\qbar$, and
$ \delta q$ denotes a variation in $L^2(\X)$.  From this calculation
we obtain the equilibrium equation $ i'(\qbar) = - \beta \psibar - \gamma$,
which we can express in the form 
\begin{equation} \label{mfeqn}
\qbar \;=\; - \Delta \psibar + r^{-2} \psibar + b 
      \;=\; f'(- \beta \psibar - \gamma) \, . 
\end{equation}
The last expression uses the fact that, since $f$ and $i$ are
conjugate convex functions, their first derivatives $f'$ and $i'$ are
inverse functions.  Thus, the statistical equilibrium model produces a
semilinear elliptic equation for the streamfunction $\psibar$ of the
most probable flow. We shall refer to (\ref{mfeqn}) as the {\em
mean-field equation}.  The predicted dependence $f'$ of the mean
potential vorticity on the mean streamfunction is determined solely by
the statistical properties of the small-scale fluctuations in the
model, since the prior distribution $\rho$ determines $f$ through
(\ref{cgf}).  With a fixed prior distribution $\rho$, the branches of
most probable, or coherent, states are parametrized by $\beta$ and
$\gamma$, which enter nonlinearly in (\ref{mfeqn}).  The mean-field
equation can possess nonunique solutions, and its solutions branches
can bifurcate.

Let us also record the second-order conditions at an equilibrium state
$\qbar$.  With $\delta \psi$ denoting the solution to $( - \Delta +
r^{-2} ) \, \delta \psi = \delta q$ under appropriate boundary
conditions, there holds
\begin{eqnarray} \label{second_order}
0 & \le & \delta^2 (\, I + \beta \, H + \gamma \, C \,)(\qbar)  \\
  & = & \int_{\X} \, \left\{ \, i''(\qbar) (\delta q)^2 \,+\,
 \beta  \left[ \left(\frac{ \d \delta \psi}{ \d x_1}\right)^2  \,+\,
    \left(\frac{ \d \delta \psi}{ \d x_2}\right)^2  \;+\;
       r^{-2} (\delta \psi)^2 \right]  \right\} \; dx \; . \nonumber
\end{eqnarray}
This condition is equivalent to the nonnegative-definiteness of the
bounded, symmetric operator $i''(\qbar)+ \beta G$ on $L^2(\X)$, where
$i''(\qbar)$ is a multiplication operator and $G$ is the Green
operator (\ref{green_operator}) .  Both of these component operators
are positive-definite.  Consequently, the second-order conditions are
automatically satisfied whenever $\beta$ is positive.  When $\beta$ is
negative, however, a critical point $\qbar$ satisfying the mean-field
equations is not an equilibrium state unless the second variation of
$I+\beta H$ is nonnegative-definite at $\qbar$.  Accordingly, the
second-order conditions are crucial in the negative temperature
regime, which is often the regime of most interest in the study of
isolated coherent structures.  Finally, we note that if the second
variation is {\em strictly} positive-definite at $\qbar$, then
$\E_{\beta,\gamma} = \{ \qbar \}$, and the equilibrium is isolated and
nondegenerate.  Conversely, the degeneracy of the second variation
signals the presence of a phase transition.

\vspace{.2in}
\noindent
{\bf 3.3 Microcanonical model.} In some respects the microcanonical
ensemble (\ref{micro_distr}) defines a more natural model than the
corresponding canonical ensemble.  From a physical point of view, the
canonical parametrization of equilibrium states by an inverse
temperature $\beta$ and a chemical potential $\gamma$ is undesirable
because the coherent structures are not maintained by contact with a
bath having these parameters.  Rather, the equilibrium states
represent organized flows on the large scales which contain the energy
$E$ and circulation $\Gamma$ and are isolated from the turbulent
fluctuations on the small-scales.  It is therefore reasonable to
parametrize such flows by $E$ and $\Gamma$.  From a mathematical
standpoint, we are compelled to study the microcanonical formulation
of the statistical equilibrium theory by virtue of the fact, which we
establish in Section 4, that the microcanonical model is not always
equivalent to the canonical model in the continuum limit.  

The following theorem characterizes the continuum limit for the
microcanonical model in terms of the coarse-grained process $\Qtilde$.

\vspace{.2in}
\noindent
{\bf Theorem 3.}  With respect to the microcanonical ensemble
$P_n^{E,\Gamma}(dQ)$, the coarse-grained process $\Qtilde$
satisfies the double limit
\begin{equation} \label{micro_ldp}
\lim_{\ntilde \rightarrow \infty} \lim_{n \rightarrow \infty} 
\frac{1}{n} \log P_n^{E,\Gamma} \{ \Qtilde \in B \} \;=\; 
                  - I^{E,\Gamma}(B) \, , 
\end{equation}
for any Borel subset $B$ of $L^2(\X)$ that is a continuity set for
$I^{E,\Gamma}$; in this formula,
\begin{equation}  \label{micro_info}
I^{E,\Gamma} (q) \; \doteq \; \left\{  \begin{array}{ll}
                                       I(q) + S(E,\Gamma) &
                \;\;\;\;\;\;   \mbox{ if } H(q)=E, \; C(q)=\Gamma  , \\
                                       + \infty  & 
                \;\;\;\;\;\;   \mbox{ otherwise, } 
                                       \end{array}  \right.
\end{equation}
where 
\begin{eqnarray} \label{micro_entropy}
S(E,\Gamma)  & \doteq &  
 - \min \, \{ \, I(q) \, : \, H(q)=E, \;  C(q)=\Gamma \, \}  \\
                    &  =  & 
 \lim_{n \rightarrow \infty} 
         \frac{1}{n} \log \Pi_n \{ \, H_n=E, \; C_n=\Gamma  \, \} \, .   \nonumber
\end{eqnarray}
\vspace{.15in}

We reiterate our earlier remark that we have taken the microcanonical
constraints to be exact equalities for the sake of clarity in the
exposition.  To obtain mathematically rigorous versions of these
results, we first replace the microcanonical constraints by the
containments $H_n \in [E - \epsilon, E + \epsilon]$ and $C_n \in [
\Gamma - \delta, \Gamma + \delta]$ with finite $\epsilon,\delta >0$,
and we then take a third limit as $\epsilon,\delta \rightarrow 0$ in
(\ref{micro_ldp}) after the limits on $n$ and $\ntilde$.

This theorem is a simplified version of a general theorem that we
formulate and prove in our companion paper \cite{EHT1}.  As in the
analysis of canonical model, the representations
(\ref{representation}) are fundamental to the proof.  With these
approximations and the basic large deviation principle
(\ref{base_ldp}) in hand, the large deviation principle
(\ref{micro_ldp}) can be deduced directly from the general results in
\cite{EHT1}.

The large deviation principle for the microcanonical ensemble involves
the {\em microcanonical information functional} $I^{E,\Gamma}$.  Among
the macrostates lying on the microcanonical manifold $H=E,C=\Gamma$,
the most probable macrostates $q$ are those at which $I^{E,\Gamma}$
achieves its minimum value of $0$. These macrostates compose the {\em
set of equilibrium states} associated with given microcanonical
parameters $E >0, \Gamma \in R$; namely,
\begin{equation} \label{micro_equil}
\E^{E,\Gamma} \; \doteq \; 
                     \{ q \in L^2(\X) \, : \, I^{E,\Gamma} (q) = 0 \, \}
                  \;=\; \arg \min \, \{ \, I \, : H = E, \; C = \Gamma  \, \}
\end{equation}
As in the canonical model, any macrostate $q$ that does not lie in the
equilibrium set has an exponentially small probability of being
observed as a coarse-grained state in the continuum limit.
Conversely, the equilibrium macrostates in $\E^{E,\Gamma}$, which
solve the constrained minimization problem with objective functional
$I$, are the overwhelmingly most probable coarse-grained states
compatible with the microcanonical constraints $H=E, \, C=\Gamma$.
Again, as in the canonical model, the existence of an equilibrium
state $\qbar$ in $\E^{E,\Gamma}$ is ensured by direct methods. Since
the equality constraint $H=E$ makes the microcanonical manifold a
nonconvex set, constrained minimizers may be nonunique, and hence
$\E^{E,\Gamma}$ may contain multiple equilibrium macrostates.

The first-order conditions for a microcanonical equilibrium $\qbar \in
\E^{E,\Gamma}$ are identical to (\ref{first_order}), except that
$\beta$ and $\gamma$ are Lagrange multipliers for the energy and
circulation constraints, respectively.  The solution triple
$(\qbar,\beta,\gamma)$ is determined, in principle, by the given
constraint pair $(E,\Gamma)$, since the multipliers are uniquely
determined by the critical point $\qbar$.  Similarly, the mean-field
equation (\ref{mfeqn}) holds without change in the microcanonical
model, except that the parameters $\beta$ and $\gamma$ appearing in it
are also unknowns.

The second-order conditions, on the other hand, are fundamentally
altered by shifting from the canonical to microcanonical formulation.
>From general principles in optimization, we know that the
nonnegativity condition (\ref{second_order}) at a constrained
minimizer $\qbar$ holds for all variations $\delta q$ that are
infinitesimally compatible with the constraints, but not necessarily
for arbitrary variations $\delta q \;$ \cite{IT,Zeidler}.  Thus, we find
that the second-order conditions appropriate to a macrostate $\qbar
\in \E^{E,\Gamma}$ are that (\ref{second_order}) holds for all $\delta
q$ satisfying the linearized side-conditions
\begin{equation} \label{tangent_var}
\delta H(\qbar) \;=\; \int_{\X} \psibar \, \delta q \, dx \;=\; 0 
\;\;\;\; \mbox{ and } \;\;\;\;
\delta C(\qbar) \;=\; \int_{\X} \, \delta q \, dx \;=\; 0 \, .  
\end{equation}
Given this characterization of the constrained minimizers of $I$
subject to $H=E$ and $C=\Gamma$, we see that set of microcanonical
equilibria is potentially larger than the corresponding set of
canonical equilibria. 

This difference between the canonical and microcanonical equilibrium
equations at second-order underlies all of our subsequent development.
Broadly speaking, it implies that families of microcanonical
equilibria are richer than corresponding families of canonical
equilibria, and that nonlinear stability criteria based on the
microcanonical formulation are finer than corresponding criteria for
the canonical formulation.

%==========================================================================
\section{Equivalence and nonequivalence}
%subsections: 
%4.1 Thermodynamic functions and equilibria 
%4.2 Illustrative example
%
We now turn our attention to the relation between the equilibrium
sets $\E_{\beta,\gamma}$ for the canonical model and the equilibrium
sets $\E^{E,\Gamma}$ for the microcanonical model.  In most
statistical equilibrium models, the canonical and microcanonical
ensembles are equivalent, in the sense that there is a one-to-one
correspondence between their equilibrium states.  For the local
mean-field models of coherent structures in turbulence, however, there
can be microcanonical equilibria that cannot be realized as canonical
equilibria.  Moreover, these equilibrium states are neither rare nor
pathological.  Rather, they are often the coherent mean flows of
greatest physical interest.  In the analysis to follow, we show how
the properties of the thermodynamic functions in the microcanonical
and canonical models determine the correspondence, or lack of
correspondence, between equilibria for these two models.

\vspace{.2in}
\noindent
{\bf 4.1 Thermodynamic functions.}  The fundamental thermodynamic
function for the microcanonical model is the value function
$S(E,\Gamma)$ in the constrained maximum entropy principle
(\ref{micro_entropy}) whose solutions constitute the equilibrium set
$\E^{E,\Gamma}$.  Similarly, the fundamental
thermodynamic function for the canonical model is the value function
$\Phi(\beta,\gamma)$ in the free maximum entropy principle
(\ref{free_energy}) whose solutions constitute the equilibrium set
$\E_{\beta,\gamma}$.  These two functions are conjugate functions in
the sense of convex analysis \cite{IT,Zeidler}; that is, they are
related by the identity
\begin{equation} \label{conjugate}
\Phi(\beta,\gamma) \;=\; \inf_{E,\Gamma} \, [ \, \beta E + \gamma \Gamma 
                            - S(E,\Gamma) \, ]
\end{equation}
The proof simply amounts to writing the free minimization in
(\ref{free_energy}) in terms of the constrained minization in
(\ref{micro_entropy}):
\begin{eqnarray*} 
\min_{q} \, [ \, I + \beta H + \gamma \Gamma \, ] &=&
\inf_{E,\Gamma} \, \min_{q} \, \{ \,  I + \beta H + \gamma \Gamma \, :
             \, H=E, \, C=\Gamma \, \} \\   &=&
\inf_{E,\Gamma} \, [ \, \beta E + \gamma \Gamma - S(E,\Gamma) \, ] \, .
\end{eqnarray*}
In other words, $\Phi = S^*$ is the Legendre-Fenchel transform of $S$.
Consequently, $\Phi$ is a concave function of $(\beta,\gamma)$, which
runs over $R^2$.  By contrast, $S$ itself is not necessarily concave.
The concave hull of $S$ is furnished by the conjugate function of
$\Phi$, namely, $\Phi^*=S^{**}$, which satisfies the inequality
\begin{equation} \label{hull}
S(E,\Gamma) \; \le \; 
\inf_{\beta,\gamma} \, 
   [ \, \beta E + \gamma \Gamma - \Phi(\beta,\gamma) \, ] 
      \;=\; S^{**} (E,\Gamma) \, . 
\end{equation}

The relation between microcanonical equilibria and canonical
equilibria depends crucially on the concavity properties of the
microcanonical entropy $S$.  Henceforth, we shall consider the
function $S$ to be defined on a domain $\A$, which we take to be the
largest open subset of $R^2$ consisting of admissible constraint pairs
$(E,\Gamma)$ for the microcanonical model; such a constraint pair is
admissible if $(E,\Gamma) = (H(q),C(q))$ for some $q \in L^2(\X)$ with
$I(q) < +\infty$.  We call this domain $\A$ the {\em admissible set}
for the microcanonical model.  Since, in general, $S$ is not a concave
function on $\A$, we introduce the subset $\C \subseteq \A$ on which
the concave hull $S^{**}$ coincides with $S$; that is, $(E,\Gamma) \in
\C$ if and only if $ S^{**}(E,\Gamma) = S(E,\Gamma)$.  There is
another equivalent definition of $\C$.  Namely, $\C$ consists of those
points $(E,\Gamma) \in \A$ for which there exists some $(\beta,\gamma)
\in R^2$ such that
\begin{equation} \label{support_plane}
S(E',\Gamma') \, \le \, 
   S(E,\Gamma) + \beta(E'-E) + \gamma (\Gamma' - \Gamma)
\end{equation}
for all $(E',\Gamma') \in \A$.  This condition means that $S$ has a
supporting plane, with normal determined by $(\beta,\gamma)$, at the
point $(E,\Gamma)$.  Such points $(E,\Gamma)$ are precisely those
points of $\A$ at which $S$ has a (nonempty) superdifferential, which
is the set of all $(\beta,\gamma)$ for which the above condition holds
\cite{IT,Zeidler}.

\vspace{.2in}
\noindent
{\bf 4.2 Microcanonical and canonical equilibrium sets.}  The set
$\C$, which we call the {\em concavity set}, plays a pivotal role in
the criteria for equivalence of ensembles.  The following theorem
gives results of this kind.

\vspace{.2in} 
\noindent
{\bf Theorem 4.} 
\newline  \noindent
(a) If $(E,\Gamma) \in \A$ belongs to $ \C$, then $\E^{E,\Gamma}
\subseteq \E_{\beta,\gamma}$ for some $(\beta,\gamma)$.  
\newline  \noindent
(b) If $(E,\Gamma) \in \A$ does not belong to $\C$, then 
$\E^{E,\Gamma} \bigcap \E_{\beta,\gamma} \,=\,
\emptyset$ for all  $(\beta,\gamma)$.  
\\

{\bf Proof.}  (a) The hypothesis means that equality holds in
(\ref{hull}) and is attained at some $(\beta,\gamma)$ for which
\[
S(E,\Gamma) \,=\, \beta E + \gamma \Gamma - \Phi(\beta,\gamma) \, .
\]
To show the claimed containment, take any $\qbar \in \E^{E,\Gamma}$
and note that $H(\qbar)=E, \, C(\qbar)=\Gamma$ and $I(\qbar) = -
S(E,\Gamma)$.  Substitution of these expressions into the above
equality yields
\begin{eqnarray*} 
I(\qbar) + \beta H(\qbar) + \gamma C(\qbar) &=&
-S(E,\Gamma) +\beta E + \gamma \Gamma \\ 
&=& \Phi(\beta,\gamma) \;=\; 
 \min_{q} \, [ \, I(q) + \beta H(q) + \gamma C(q) \,] \, ,
\end{eqnarray*}
using (\ref{free_energy}).  Since $\E_{\beta,\gamma}$ consists of the
minimizers of $I + \beta H + \gamma C$, it follows that $ \qbar \in
\E_{\beta,\gamma}$.  This completes the proof of (a).

(b) A complementary argument to that used in (a) applies.  
Now, the hypothesis means that, for all  $(\beta,\gamma)$,
\[
S(E,\Gamma) \,<\, \beta E + \gamma \Gamma - \Phi(\beta,\gamma) \, .
\]
Then,  any $\qbar \in \E^{E,\Gamma}$ satisfies
\begin{eqnarray*} 
I(\qbar) + \beta H(\qbar) + \gamma C(\qbar) &=&
-S(E,\Gamma) +\beta E + \gamma \Gamma \\ 
&>& \Phi(\beta,\gamma) \;=\; 
 \min_{q} \, [ \, I(q) + \beta H(q) + \gamma C(q) \,] \, .
\end{eqnarray*}
Thus, $\qbar$ does not minimize $I + \beta H + \gamma C$, and 
hence does not belong to $\E_{\beta,\gamma}$.  Since
$(\beta,\gamma)$ is arbitrary, this completes the proof of (b).

>From Theorem 4 we see that, for constraint pairs in $\C$, the
microcanonical equilibria are contained in a corresponding canonical
equilibrium set, while, for constraint pairs in $\A \backslash \C$, the
microcanonical equilibria are not contained in any canonical
equilibrium set.  Consequently, whenever $\C \neq \A$ the canonical
equilibria do not exhaust the admissible microcanonical constraint
pairs, and there are microcanonical equilibria that are not realized
by any canonical equilibria.  On the other hand, all canonical
equilibria are contained in some microcanonical equilibrium set, and
$\C$ is exhausted by the constraint pairs realized by all canonical
equilibria.  These further results are given in the following theorem.   

\vspace{.2in}
\noindent
{\bf Theorem 5.} 
\newline  \noindent
(a) The concavity set $\C$ consists of all constraint pairs realized
by the canonical equilibria; that is,  
\begin{equation} \label{concavity_set}
\C \;=\; \bigcup  \, 
       \{ \, ( H, C ) ( \E_{\beta,\gamma} ) \, : \, 
              ( \beta, \gamma ) \in R^2 \,  \} \, .
\end{equation}
(b) Each canonical equilibrium set  $\E_{\beta,\gamma}$ consists
of all microcanonical equilibria whose constraint pairs are realized
by  $\E_{\beta,\gamma}$; that is, for any  $(\beta, \gamma)$, 
\begin{equation} \label{can_realized}
 \E_{\beta,\gamma} \;=\;  \bigcup \, 
  \{ \, \E^{E,\Gamma} \, : 
              \; (E,\Gamma) \in (H,C) ( \E_{\beta,\gamma}) \, \} \, .   
\end{equation}

\vspace{.2in} {\bf Proof.} 
(a) The containment of $\C$ in the union is immediate
from Theorem 4a.  To show the opposite containment we argue by
contradiction, supposing that for some $(\beta,\gamma)$ and some
$\qbar \in \E_{\beta,\gamma}$, $(E,\Gamma) = (H(\qbar),C(\qbar)) 
\in \A \backslash \C$.  Then, we find that 
\begin{eqnarray*} 
S(E,\Gamma) & < & \beta E + \gamma \Gamma - \Phi(\beta,\gamma)  \\ 
            & = & -I(\qbar) \; \le \; S(E,\Gamma) \, , 
\end{eqnarray*}
using (\ref{free_energy}) and (\ref{micro_entropy}) as in the
proof of Theorem 4b.  We thus obtain the desired contradiction.
This completes the proof of (a).  

(b) The containment of $\E_{\beta,\gamma}$ in
the union is straightforward.  Let $\qbar \in \E_{\beta,\gamma}$, and
set $E=H(\qbar)$ and $\Gamma=C(\qbar)$. Then, $I(\qbar) + \beta E +
\gamma \Gamma \, \le \, I(q) + \beta H(q) + \gamma C(q)$ for all $q$.
For those $q$ which satisfy the constraints $H(q)=E, \, C(q)=\Gamma$,
we therefore find that $I(\qbar) \le I(q)$.  Hence, $\qbar \in
\E^{E,\Gamma}$.

The opposite containment is also straightforward.  If $E=H(\tilde{q})$
and $\Gamma=C(\tilde{q})$ for some $\tilde{q} \in \E_{\beta,\gamma}$,
then for any $\qbar \in \E^{E,\Gamma}$, we have $I(\qbar) \le
I(\tilde{q})$.  Since $\tilde{q} \in \E_{\beta,\gamma}$, we obtain
\begin{eqnarray*} 
\min_{q} \, [ \, I(q) + \beta H(q) + \gamma C(q) \, ]
&=& I(\tilde{q}) + \beta E + \gamma \Gamma   \\
& \ge &  I(\qbar) + \beta H(\qbar) + \gamma C(\qbar) \, .
\end{eqnarray*}
Hence, $\qbar \in \E_{\beta,\gamma}$.  This completes the proof of (b).

Theorems 4 and 5 allow us to classify the microcanonical constraint
parameters $(E,\Gamma)$ according to whether or not equivalence of
ensembles holds for those parameters.  In fact, the admissible set
$\A$ can be decomposed into three disjoint sets, where (1) there is a
one-to-one correspondence between microcanonical and canonical
equilibria, (2) there is a many-to-one correspondence from
microcanonical equlibria to canonical equilibria, and (3) there is no
correspondence.  In order to simplify the precise statement of this
result, let us assume that the microcanonical entropy $S(E,\Gamma)$ is
differentiable on its domain $\A$.  Then, for each microcanonical
parameter $(E,\Gamma) \in \A$ there is a corresponding canonical
parameter $(\beta,\gamma)$ determined locally by
\[
\beta = \frac{\d S}{\d E} \, , \;\;\;\;\;\;\;\;
\gamma = \frac{\d S}{\d \Gamma} \, . 
\] 
Under this assumption, we have the following classification.  

\vspace{.15in}
\noindent 
{\em 1. Full equivalence.} If $(E,\Gamma)$ belongs to $\C$ and there
is a unique point of contact between $S$ and its supporting plane at
$(E,\Gamma)$, then $\E^{E,\Gamma}$ coincides with $\E_{\beta,\gamma}$

\noindent
{\em 2. Partial equivalence.} If $(E,\Gamma)$ belongs to $\C$ but
there is more than one point of contact between $S$ and its supporting
plane at $(E,\Gamma)$, then $\E^{E,\Gamma}$ is a strict subset of
$\E_{\beta,\gamma}$.  Moreover, $\E_{\beta,\gamma}$ contains all
those $\E^{E',\Gamma'}$ for which $(E',\Gamma')$ is also a point of
contact.

\noindent
{\em 3. Nonequivalence.}  If $(E,\Gamma)$ does not belonging to $\C$,
then $\E^{E,\Gamma}$ is disjoint from $\E_{\beta,\gamma}$.  In fact,
$\E^{E,\Gamma}$ is disjoint from all canonical equilibrium sets.
\vspace{.15in}

The proofs of these results can be constructed easily using the same
techniques as in the proofs of Theorems 4 and 5.  We therefore leave
the necessary demonstrations to the reader.  We give a complete
discussion of these results in a more general setting in our paper
\cite{EHT1}, where we state and prove the corresponding results
without the simplifying assumption that $S$ is differentiable.
Experience with numerical solutions of these variational problems of
this kind, however, strongly suggests that the differentiability
assumption is essentially always satisfied.  These computations
also show that the parameter regime of nonequivalence can be quite
wide and can contain many physically interesting equilibrium flows.
In Section 6, we present a computed example that illustrates this
behavior.

%==========================================================================
\section{Nonlinear stability}
%subsections: 
%5.1 Arnold stability
%5.2 Refined stability
%
In either the canonical or the microcanonical model, the equilibrium
macrostates determine steady mean flows that are the most probable
flows compatible with the given parameters of the model.  This
statistical property of the mean flows can be interpreted as a
stability property in a weak sense. That is, while the underlying
ergodic dynamics continually produces unsteady perturbations in the
microstate, the coarse-grained macrostate remains near the mean flow
with very high probability.  In other words, the construction of the
steady mean flows as statistical equilibrium macrostates guarantees
that they are stable with respect to perturbations on the microscopic
scales.  We now inquire whether these steady mean states are also
stable in a strong sense with respect to macroscopic perturbations.
Precisely, we investigate the evolution under ideal dynamics of any
perturbed macroscopic state $q(t)$ that initially lies within a small,
finite distance $\| q(0) - \qbar \|$ in $L^2(\X)$ of an equilibrium
macrostate $\qbar$.

In the canonical model, we find that the most probable state $\qbar$
for any $\beta$ and $\gamma$ satisfies the celebrated Arnold stability
criteria, the canonical information functional $I_{\beta,\gamma}$
being the required Lyapunov functional.  We collect these results in
Subsection 5.1.  In the microcanonical model, on the other hand, we
encounter a gap in the classical stability criteria in the sense that
there are microcanonical equilibria which are stable, but for which
$I_{\beta,\gamma}$ does not satisfy the conditions needed in the
Lyapunov stability argument.  In Subsection 5.2, we therefore devise a
more refined argument based on a penalization of this functional and
thereby fill the gap in the known stability theorems.

\vspace{.2in} \noindent
{\bf 5.1 Arnold stability theorems.} The equilibria for the canonical
model correspond to steady flows that satisfy the nonlinear stability
criteria of Arnold \cite{Arnold,MP}.  In this subsection we
reformulate these classical results in the context of the statistical
equilibrium theory.

Throughout this discussion we assume that for given values of the
canonical parameters $\beta$ and $\gamma$, the equilibrium state
$\qbar \in \E_{\beta,\gamma}$ is an isolated, nondegenerate minimizer
of canonical information functional $I_{\beta,\gamma}$; otherwise, the
stability of a single equilibrium state $\qbar$ cannot be expected.
The fact that $\qbar$ is a minimizer of $I_{\beta,\gamma}$ over
$L^2(\X)$ guarantees that the second variation $\delta^2
I_{\beta,\gamma}(\qbar) $ appearing in (\ref{second_order}) is
nonnegative definite.  A sufficient condition for $\qbar$ to be a
nondegenerate minimizer is that $\delta^2 I_{\beta,\gamma}(\qbar)$ be
strictly positive definite.  More precisely, we say that $\qbar$ is an
{\em nondegenerate} canonical equilibrium state if
\begin{equation} \label{dtwo_i_lower}
\mu \int_{\X} \, (\delta q)^2 \, dx \; \le \; 
    \delta^2 \, I_{\beta,\gamma} (\qbar)
\end{equation}
for all variations $\delta q \in L^2(\X)$, with a positive constant
$\mu$ independent of $\delta q$.  The optimal constant $\mu$ in
(\ref{dtwo_i_lower}) is the smallest eigenvalue of the operator
$i''(\qbar) + \beta G$, where $G$ is the Green operator
(\ref{green_operator}).  This fact is immediate from the identity
\[
\delta^2 \, I_{\beta,\gamma}(\qbar) \;=\; 
    \delta^2 \, ( \, I + \beta H + \gamma \Gamma \, ) (\qbar) \;=\;
       \int_{\X} [ \, i''(\qbar) (\delta q )^2 \,+\, \beta \, 
                  \delta q \, G \,  \delta q \, ] \, dx \; .  
\] 
An upper bound that complements the lower bound (\ref{dtwo_i_lower})
also holds, namely, 
\begin{equation} \label{dtwo_i_upper}
\delta^2 \, I_{\beta,\gamma}(\qbar)  \; \le \; 
   \nu \int_{\X} \, (\delta q)^2 \, dx \; .
\end{equation}
In this upper bound it suffices to take the constant $\nu = \max
i''(\qbar) \, + \, |\beta|/ \lambda_1$, where $\lambda_1>0$ is the
smallest eigenvalue of $-\Delta + r^{-2}$; the required bound on
$i''(\qbar) \,=\, 1/f''( - \beta \psibar - \gamma)$ follows easily
from the fact that $f''(\eta)$ equals the variance of the distribution
$z(\eta)^{-1} \, e^{\eta y} \rho(dy)$, which is bounded below by a
positive constant uniformly for $\eta$ in a bounded interval.

The nonlinear stability result for the canonical model is summarized
in the following theorem.

\vspace{.2in} 
\noindent
{\bf Theorem 6.} If $\qbar \in \E_{\beta,\gamma}$ is a nondegenerate
canonical equilibrium state, then the corresponding steady flow is
stable; specifically, if $q(t)$ denotes the solution to (\ref{qg}) and
if $\| q(0) - \qbar \|$ is sufficiently small, then for all time $t
>0$
\[
\| q(t) - \qbar \| \; \le \; c \, \| q(0) - \qbar \|
\]
for some finite constant $c$.  
\vspace{.2in} 

{\bf Proof.} The proof relies on the fact that $I_{\beta,\gamma}$ is a
conserved quantity for the dynamics (\ref{qg}).  The conservation of
$H$ and $C$ is immediate, since they are rugged invariants.  The
information functional $I$ is also an invariant under the ideal
dynamics that governs $q(t)$, since it coincides with a certain
generalized enstrophy integral (\ref{enstrophy}) under the
identification $a=i$.  We claim that the invariant $I_{\beta,\gamma}$
satisfies
\[
\frac{\mu}{2} \| q - \qbar \|^2 \; \le \; I_{\beta,\gamma}(q)  
                 \; \le \;  2 \nu \| q - \qbar \|^2 \, .
\]
for all $q$ in a small $L^2$-neigborhood of $\qbar$. These estimates
follow from the upper and lower bounds on the second variation
$\delta^2 I_{\beta,\gamma} (\qbar)$ given in (\ref{dtwo_i_lower}) and
(\ref{dtwo_i_lower}), in view of the fact that $ I_{\beta,\gamma}
(\qbar) = 0$ and $ \delta I_{\beta,\gamma} (\qbar) = 0$.  The
derivation makes use of a standard estimation of the remainder terms
in the second-order Taylor expansion of the smooth functional
$I_{\beta,\gamma}$ about $\qbar$.  Then, the usual Lyapunov argument
yields
\begin{eqnarray*}
\frac{\mu}{2} \| q(t) - \qbar \|^2  \, &\le& \,  
           I_{\beta,\gamma}(q(t)) \,  \\ 
               &=& \,  I_{\beta,\gamma}(q(0))  \; \le \;
                 2 \nu \| q(0) - \qbar \|^2 
\end{eqnarray*}
for all $t>0$, thereby proving the theorem.   

We remark that this proof of Lyapunov stability requires only that $
I_{\beta,\gamma}(q(t)) \le I_{\beta,\gamma}(q(0))$ for $t>0$.  This
observation allows us to make a connection with the Turkington model
\cite{Turkington}, which is based on an argument that only
inequalities on convex generalized enstrophies constrain the ideal
dynamics.  Even though $I$ is treated as a fragile invariant in that
framework, the nonlinear stability of $\qbar$ remains valid, since $H$
and $C$ are rugged invariants.

In the context of the statistical equilibrium theory, the classical
stability criteria amount to sufficient conditions for the
nondegeneracy of the minimizer $\qbar$.  For positive temperature
states ($\beta \ge 0$), the so-called first Arnold theorem applies,
while for negative temperature states ($\beta <0$), the so-called
second Arnold theorem applies \cite{Arnold,MP}.  In either case the
sufficient condition for stability is that the bounded, symmetric
operator $ i''(\qbar) + \beta G$ be positive definite.  This form of
the stability condition can be translated into the familiar form used
in deterministic studies of steady flows by means of the formula
\[
\frac{d \qbar}{d \psibar} \;=\; - \frac{\beta}{i''(\qbar)} \, ,
\]
which follows from the mean-field equation (\ref{mfeqn}) and the fact
that $f'$ and $i'$ are inverse functions.  In this form, the first
Arnold theorem applies when $d \qbar / d \psibar \, < \, 0$, while the
second Arnold theorem applies when $ 0 \, < \, d \qbar / d \psibar \,
< \lambda_1 \,$, where $\lambda_1$ is the smallest eigenvalue of
$-\Delta + r^{-2}$.  If a deterministic steady flow corresponding to a
potential vorticity field $\qbar$ is submitted to these stability
criteria, there often are instances when neither the first nor the
second theorem applies; these steady flows correspond to critical
points for $I_{\beta,\gamma}$ at which its second-variation is
negative in some direction.  By constrast, any nondegenerate canonical
equilibrium state $\qbar$ satisfies these criteria, the first when
$\beta \ge 0$ and the second when $\beta < 0$.  Thus, apart from
degeneracies such as occur at phase transitions, the statistical
equilibrium theory always produces mean flows that are both steady and
stable.

\vspace{.2in}
\noindent
{\bf 5.2 Refined stability theorems.}  When a microcanonical
equilibrium $\qbar \in \E^{E,\Gamma}$ does not lie in any canonical
equilibrium set $\E_{\beta,\gamma}$, the stability results of the
preceding subsection do not apply.  Nevertheless, every nondegenerate
equilibrium state for the microcanonical model determines a stable
flow, as we now show by giving a more refined nonlinear stability
analysis.

Again, we assume that $\qbar \in \E^{E,\Gamma}$ is the isolated,
nondegenerate minimizer of the microcanonical information $I$ at given
microcanonical constraint values $E$ and $\Gamma$.  In the
microcanonical model, however, the second-order conditions at a
constrained minimizer $\qbar$ are subject to side-conditions on
$\delta q$, which are the linearization of the constraints $H=E, \,
C=\Gamma$.  Precisely, we say that $\qbar$ is a {\em nondegenerate}
microcanonical equilibrium state if (\ref{dtwo_i_lower}) holds for all
$\delta q \in L^2(\X)$ that satisfy the linearized constraints
(\ref{tangent_var}), with a positive constant $\mu$ independent of
these $\delta q$.  The complementary upper bound (\ref{dtwo_i_upper})
also holds at the microcanonical equilibrium $\qbar$, with a constant
$\nu$ determined as in the canonical model; in fact, the upper bound
also holds for $\delta q$ not satisfying the side-conditions
(\ref{tangent_var}).

Our strategy for proving the stability of $\qbar$ is to construct a
Lyapunov functional in the form
\begin{eqnarray} \label{lyapunov}
L^{E,\Gamma}_{\sigma,\tau}(q) \; & \doteq &  \;
I(q) \, + \, S(E,\Gamma) \; + \;
\beta \, [ H(q) -E ] \, + \, \gamma \, [ C(q) - \Gamma ] \;
\\ & & \;\;\;\;\;\;\;\;\;\;\;\;\;\;\; + \;
  \frac{\sigma}{2}  \, [ H(q) -E ]^2 \, + \, 
    \frac{\tau}{2} \, [ C(q) - \Gamma ]^2 \, ,
\nonumber
\end{eqnarray}
where $\beta$ and $\gamma$ are the Lagrange multipliers for the energy
and circulations constraints, respectively, and $\sigma$ and $\tau$
are sufficiently large positive constants.  The terms in
(\ref{lyapunov}) scaled by $\sigma$ and $\tau$ penalize departures
from the microcanonical constraints and thereby capture the
microcanonical conditioning in the Lyapunov functional.  Moreover,
these terms do not change the value of the Lyapunov functional or its
first variation at $\qbar$, which are
\[
L^{E,\Gamma}_{\sigma,\tau}(\qbar) \;=\; 0 \, , \;\;\;\;\;\;
\delta L^{E,\Gamma}_{\sigma,\tau}(\qbar) \;=\; 
  \delta ( \, I + \beta \, H + \gamma \, C \,) (\qbar) \;=\; 0 \, . 
\]
For this reason, it is possible to choose finite constants $\sigma$
and $\tau$ so that $L^{E,\Gamma}_{\sigma,\tau}$ has a nondegenerate,
unconstrained minimum at $\qbar$.  In this sense
$L^{E,\Gamma}_{\sigma,\tau}$ is identical to the ``augmented 
Lagrangian'' often used in numerical methods of constrained
optimization \cite{Bertsekas,Minoux}.

In the case of full equivalence, when the concavity condition
(\ref{support_plane}) holds as a strict inequality for all
$(E',\Gamma') \neq (E,\Gamma)$, the penalizing terms are unnecessary,
because $L^{E,\Gamma}_{0,0}$ coincides with $I_{\beta,\gamma}$ and
hence furnishes a Lyapunov functional at $\qbar$.  Indeed, the
argument used to prove part (a) of Theorem 4 applies to this
situation, and guarantees that $L^{E,\Gamma}_{0,0}(q) >
L^{E,\Gamma}_{0,0}(\qbar) =0 $ for all $q \neq \qbar$.  In the cases
of nonequivalence or partial equivalence, however, when the
microcanonical equilibrium $\qbar$ may not be contained in the
corresponding canonical equilibrium set, $L^{E,\Gamma}_{0,0}$ may not
be a Lyapunov functional at $\qbar$. In those cases, $\delta^2
L^{E,\Gamma}_{0,0}$ may be negative for variations $\delta q$ that are
not tangential to the constraint manifold $H=E, \, C=\Gamma$.  In
general, it is therefore necessary to include penalization parameters
$\sigma$ and $\tau$ so that 
$\delta^2 L^{E,\Gamma}_{\sigma,\tau}(\qbar)$ is positive definite.  
An explicit calculation of this second-variation, namely,
\begin{equation}  \label{secondvar_l} 
\delta^2 L^{E,\Gamma}_{\sigma,\tau}(\qbar) \;=\;
 \delta^2 ( \, I + \beta H + \gamma C \, ) (\qbar)
  \;+\; \sigma \left\{ \int_{\X} \, \psibar \delta q \, dx \right\}^2 
  \;+\; \tau \left\{ \int_{\X} \, \delta q \, dx \right\}^2 \, ,
\end{equation}
suggests that it is indeed positive definite on arbitrary variations
$\delta q$ when $\sigma$ and $\tau$ are sufficiently large.

The nonlinear stability result for the microcanonical model is the
content of the following theorem.   

\vspace{.2in} 
\noindent
{\bf Theorem 7.} If $\qbar \in \E^{E,\Gamma}$ is a nondegenerate
microcanonical equilibrium state, then the corresponding steady flow
is stable; specifically, if $q(t)$ denotes the solution to (\ref{qg})
and if $\| q(0) - \qbar \|$ is sufficiently small, then for all time
$t >0$
\[
\| q(t) - \qbar \| \; \le \; c \,  \| q(0) - \qbar \|
\]
for some finite constant $c$.  
\vspace{.2in} 

{\bf Proof.} The crux of the proof is to demonstrate that the second
variation of $L^{E,\Gamma}_{\sigma,\tau}(\qbar)$ is strictly
positive definite when $\sigma$ and $\tau$ as fixed large enough. This
analysis makes use of the bilinear form associated with the
operator $i''(\qbar) + \beta G$, which we denote by
\begin{equation} \label{bilinear}
D_2 (z_1,z_2) \;\doteq\; \int_{\X} \left[ \, i''(\qbar)\, z_1  z_2 \,+\,
                       \beta \, z_1 G z_2 \, \right] \, dx \, 
                 \;\;\;\;\;\;\;\; (\, z_1, z_2 \in L^2(\X)\,) \, . 
\end{equation} 
>From the identity in (\ref{second_order}), it is clear that 
$D_2(\delta q,\delta q) = \delta^2 (I + \beta H + \gamma C)(\qbar)$.
Also, we let $(z_1,z_2)=\int z_1 z_2 \, dx$ denote the inner product on
$L^2(\X)$.  

We decompose any variation $\delta q \in L^2(\X)$ into a part $\delta
q^{\parallel}$ tangent to the microcanonical manifold at $\qbar$, and
a part $\delta q^{\perp}$ orthogonal to it; that is,
\[
\delta q \, =\, \delta q^{\parallel} \, + \, \delta q^{\perp} \, ,
\]
where $\delta q^{\perp} = \xi \psibar + \eta 1 $ for some $\xi,\eta
\in R$, and $(\delta q^{\parallel},\psibar)=0, \, (\delta
q^{\parallel},1)=0$.  It is easy to verify that the functions
$\psibar$ and $1$ are linearly independent, given that $E \neq 0$ in
the microcanonical energy constraint.  Thus, the components $\xi$ and
$\eta$ are uniquely determined by $\delta q$, in that they solve the
associated normal equations.  A straightforward analysis then shows
that the inequality
\[
\frac{(\psibar, \delta q)^2}{\| \psibar \|^2} \; + \; 
\frac{(1, \delta q)^2}{\| 1 \|^2} \; = \; 
\frac{(\psibar, \delta q^{\perp})^2}{\| \psibar \|^2} \; + \; 
\frac{(1, \delta q^{\perp})^2}{\| 1 \|^2} \; \ge \; 
\theta  \| \delta q^{\perp} \|^2 
\]
holds for a positive constant $\theta$ depending on the angle between
$\psibar$ and $1$ in $L^2(\X)$.

We now substitute this decomposition into (\ref{secondvar_l}) and
analyze the resulting terms:
\begin{eqnarray} \label{allterms}
\delta^2 L^{E,\Gamma}_{\sigma,\tau}(\qbar) \;&=&\;
       D_2(\delta q^{\parallel}, \delta q^{\parallel}) \, + \,
      2 \,  D_2(\delta q^{\parallel}, \delta q^{\perp}) \, + \,
       D_2(\delta q^{\perp}, \delta q^{\perp}) \\
                                      & & \;\;\;\;\;\;\;\;\;\;\;\;
 + \; \sigma(\psibar,\delta q^{\perp})^2 \,+\, 
                \tau(1,\delta q^{\perp})^2 \, . \nonumber
\end{eqnarray}
The nondegeneracy hypothesis ensures that 
\begin{equation} \label{term_one}
 D_2(\delta q^{\parallel}, \delta q^{\parallel}) \; \ge \; 
      \mu \| \delta q^{\parallel} \|^2 \, .
\end{equation}
On the other hand, the upper bound (\ref{dtwo_i_upper}) gives
\begin{equation} \label{term_two}
 | D_2(\delta q^{\perp}, \delta q^{\perp}) | \; \le \; 
      \nu \| \delta q^{\perp} \|^2 \, .
\end{equation}
In a similar fashion the cross term is estimated by means of the
Cauchy inequality, giving
\begin{equation} \label{term_three}
 2 \, | D_2(\delta q^{\parallel}, \delta q^{\perp}) | \; \le \; 
      \nu \epsilon \, \| \delta q^{\parallel} \|^2 \,+\,
       \frac{\nu}{\epsilon} \, \| \delta q^{\perp} \|^2 \, ,
\end{equation}
for any $\epsilon >0$.  When we use (\ref{term_one}), (\ref{term_two}),
and (\ref{term_three}) to estimate the various terms in (\ref{allterms}), we
obtain the following lower bound:
\begin{eqnarray*}
\delta^2 L^{E,\Gamma}_{\sigma,\tau} (\qbar) \; &\ge& \;
\mu  \| \delta q^{\parallel} \|^2  \,-\, 
 \nu \epsilon \, \| \delta q^{\parallel} \|^2 \,-\,
  \frac{\nu}{\epsilon} \, \| \delta q^{\perp} \|^2 \,-\,
   \nu \, \| \delta q^{\perp} \|^2  \\
                                   & & \;\;\;\;\;\;\;\;
+ \; \sigma(\psibar,\delta q^{\perp})^2 \,+\, 
                 \tau(1,\delta q^{\perp})^2 \, .
\end{eqnarray*}  
We therefore choose $\epsilon = \mu/2\nu$ to make the terms in $
\delta q^{\parallel}$ definite.  Then, in order to make the terms in $
\delta q^{\perp}$ definite, we seek $\sigma$ and $\tau$ so that
\[
 \sigma(\psibar,\delta q^{\perp})^2 \,+\, \tau(1,\delta q^{\perp})^2
\; \ge \; \left[ \frac{\mu}{2} \,+\, 
              \frac{\nu}{\epsilon} \,+\, \nu \right] \,
          \| \delta q^{\perp} \|^2 \, .
\]
It suffices to set these penalization parameters so that $\sigma \,
\theta \| \psibar \|^2$ and $\tau \, \theta \| 1 \|^2$ equal the
common value $\mu/2 + \nu/\epsilon + \nu$.  With this choice, we
obtain the desired lower bound:
\begin{equation} \label{posdef_l}
\delta^2 L^{E,\Gamma}_{\sigma,\tau} (\qbar) \; \ge \;
\frac{\mu}{2} \,  \| \delta q^{\parallel} \|^2 \;+\; 
\frac{\mu}{2} \,  \| \delta q^{\perp} \|^2 \;=\; 
\frac{\mu}{2} \,  \| \delta q \|^2 \, .   
\end{equation}

Thus, $\delta^2 L^{E,\Gamma}_{\sigma,\tau}(\qbar)$ is strictly
positive definite, and hence we conclude that for all $q$ in a
sufficiently small neigborhood of $\qbar$,
\[
\tilde{\mu} \| q - \qbar \|^2 \; \le \; 
        L^{E,\Gamma}_{\sigma,\tau}(q)  
                 \; \le \;  \tilde{\nu} \| q - \qbar \|^2 \, , 
\]
for some $0 < \tilde{\mu} < \tilde{\nu} < \infty$.  The usual Lyapunov
stability argument therefore ensues, since
$L^{E,\Gamma}_{\sigma,\tau}$ is a conserved quantity for the
dynamics.  Thus, the proof of the theorem is complete.

As in the canonical model, we note that this Lyapunov stability
argument remains valid when the objective functional $I$ is treated as
a fragile invariant, since constraint functionals $H$ and $C$ are 
rugged invariants.

We conclude this discussion of stability with some remarks about the
role of the penalization in the Lyapunov functional
$L^{E,\Gamma}_{\sigma,\tau}$ and its relation to the microcanonical
entropy $S(E,\Gamma)$.  For the sake of definiteness, let us suppose
that $S(E,\Gamma)$ is smooth ($C^2$) on its domain $\A$, and let us
consider a constraint pair $(E,\Gamma)$ that does not belong to $\C$,
the concavity set.  Then, according to the results in Section 4, the
microcanonical equilibrium macrostate $\qbar$ for $(E,\Gamma)$ does
not belong to any canonical equilibrium set, and the tangent plane to
$S$ at $(E,\Gamma)$ is not a supporting plane, meaning that
(\ref{support_plane}) is violated for some $(E',\Gamma')$.
Nevertheless, it is possible to choose constants $\sigma$ and $\tau$
so that they define a supporting paraboloid to $S$ at $(E,\Gamma)$, in
the sense that
\[
S(E',\Gamma') \; \le \; S(E,\Gamma) 
           + \beta (E'-E) + \gamma (\Gamma'-\Gamma)
             + \frac{\sigma}{2}  (E'-E)^2 
               + \frac{\tau}{2}  (\Gamma'-\Gamma)^2
\]
for all $(E',\Gamma')$ in $\A$, with equality only when $(E',\Gamma')=
(E,\Gamma)$.  It follows that $L^{E,\Gamma}_{\sigma,\tau}(q) >
L^{E,\Gamma}_{\sigma,\tau}(\qbar) = 0$ for all $q \neq \qbar$, by an
argument analogous to that used in the proof of Theorem 4.  Thus, we
see that the minimal choice of the penalization constants $\sigma$ and
$\tau$ is determined by the condition that, at least locally, the
corresponding paraboloid lies above the function $S$ and contacts it
only at $(E,\Gamma)$.

%==========================================================================
\section{Numerical example}

%\vspace{.2in}
\noindent
{\bf 6.1 Barotropic flow over topography in a zonal channel.}  For the
purposes of illustrating the general results obtained in Sections 4
and 5, we now present a family of computed solutions to the
microcanonical variational principle for a particular choice of
domain, topography and prior distribution.  We especially focus on the
shape of the $S(E,\Gamma)$ surface, since it determines whether the
corresponding canonical model is equivalent and whether the Arnold
stability criteria apply to the equilibrium states.  In view of our
results in Section 4 showing that all canonical equilibria are
included among the microcanonical equilibria, there is no need to
implement a solver for the corresponding canonical variational
principle.

We take the domain to be a unit square $\X = \{ \, -0.5 < x_1 < 0.5,
\; -0.5 < x_2 < 0.5 \, \} $, which represents a normalized zonal
channel.  For the topography term $b$ in the potential vorticity
expression (\ref{pv}), we choose a simple sinusoid, $b=b(x_2)= B_2
\sin ( 2 \pi x_2 )$.  This topography is zonal, being independent of
$x_1$, and consists of the second harmonic with respect to $x_2$.

Such a zonal domain and topography can be viewed as an idealized and
simplified model of a zone-belt domain in a Jovian atmosphere
\cite{IC,Dowling,Marcus}.  In the 1-1/2-layer model, $b$ is the
effective topography that results from an underlying steady mean flow
in a deep lower layer. The domain is composed of a zone, where $b$ is
positive, and a belt, where $b$ is negative.  If the amplitude $B_2$
of the topography is large enough, one expects that the mean flow in
the shallow upper layer will a shear flow $v= (v_1(x_2),0)$, and that
it will tend to be anticyclonic (negative vorticity) in the zone and
cyclonic (positive vorticity) in the belt. In our computations of most
probable flows we set $B_2=1$, and we find that they are zonal shear
flows with the expected topography-induced tendencies.

We illustrate the effect of a large or a small radius of deformation
by choosing the representative values $r=\infty$ or $r=0.2$.  The
small deformation radius regime is the one relevant to a Jovian
atmosphere \cite{Marcus}.   

With these choices of the geometrical parameters, the formulation of
the model problem is complete once we specify a prior distribution
$\rho$, which determines the probabilistic structure of the
small-scale potential vorticity $Q$.  We select a family of
gamma distributions $\rho_{\epsilon}(dy)$ with 
mean, variance and skewness normalized as follows:
\[
\int y \rho_{\epsilon}(dy) \;=\; 0 , \;\;\;\;
\int y^2 \rho_{\epsilon}(dy) \;=\; 1 , \;\;\;\;
\int y^3 \rho_{\epsilon}(dy) \;=\; 2 \epsilon ;  
\]
the variable $y$ runs through the range of $Q$.  For small $\epsilon$,
these distributions are close to the standard normal distribution,
which they approach in the limit as $\epsilon$ goes to zero.  For
positive $\epsilon$, they are supported on the interval
$-\epsilon^{-1} \le y < + \infty$, and they have an exponential tail
in the positive $y$-direction.  They are defined explicitly by the
probability density
\[
\rho_{\epsilon}(dy) \;=\; \frac{1}{\epsilon \Gamma ( \epsilon^{-2} ) }
  \exp ( \epsilon^{-2} 
          [ \log (1+\epsilon y) \,-\, (1+\epsilon y) ] ) \, dy \, .
\]

The family of prior distributions $\rho_{\epsilon}(dy)$ have the
virtue that their cumulant generating functions $f_{\epsilon}(\eta)$,
which are defined by (\ref{cgf}), can be calculated explicitly;
namely,
\begin{eqnarray} \label{cgf_eps} 
f_{\epsilon}(\eta) &=& - \epsilon^{-1} \eta \,-\, 
                           \epsilon^{-2} \log (1 - \epsilon \eta ) \\
          &=& \eta^2 / 2 \;+\; \epsilon \eta^3 / 3 \; + \;
               O(\epsilon^2) \, . \nonumber
\end{eqnarray}
The associated information functional $I_{\epsilon}$, defined in
(\ref{info}), is then determined by the conjugate function
$i_{\epsilon}(y)$ to $f_{\epsilon}(\eta)$; namely,
\[
i_{\epsilon}(y) \;=\; \epsilon^{-1} y \,-\, 
                           \epsilon^{-2} \log (1 + \epsilon y ) \, .
\]
The relevant properties of the convex function $i_{\epsilon}(y)$ are
easily seen from its second derivative, 
$i_{\epsilon}''(y) = (1 + \epsilon y)^{-2}$.  

The mean-field equation (\ref{mfeqn}) corresponding to this choice
of prior distribution is
\begin{eqnarray} \label{mfeqn_eps}
\qbar \;=\; - \Delta \psibar \,+\, r^{-2} \psibar \,+\, 
               B_2 \sin 2 \pi x_2
      &=&  \epsilon^{-1} 
 \left( [1-\epsilon (- \beta \psibar - \gamma )]^{-1} \,-\, 1 \right) \\
      &=& (- \beta \psibar - \gamma ) \;+\; 
             \epsilon (- \beta \psibar - \gamma )^2 \;+\; 
                 O(\epsilon^2) \, . \nonumber
\end{eqnarray}
>From the above expansion it is evident that $\epsilon$ determines the
magnitude of the principal nonlinear term in this equation.  When
$\epsilon = 0$, the models resemble the so-called energy-enstrophy
theory, in which the statistical equilibrium distributions are
Gaussian and the mean-field equations are linear
\cite{Kraichnan,SHH,CF,MH}.  For the sake of definiteness, we fix
$\epsilon=0.1$ in the computations to follow.  It is worth noting that
$\epsilon$ links the skewness of the prior distribution to the
nonlinearity of the mean-field equation.

While many reasonable choices of prior distribution suffice for the
purposes of the present example, the relation between the potential
vorticity and streamfunction in (\ref{mfeqn_eps}) is distinguished by
the fact that it agrees with the form of the relation inferred by an
analysis of observed zonal winds on Jupiter \cite{Dowling}.  We note
however that this physically interesting prior distribution violates
the growth condition (\ref{prior_growth}) assumed for simplicity in
our discussion of the general theory.  Nevertheless, all of the key
results described in the preceding sections remain valid for this
prior distribution, although their proofs are somewhat more involved.
In particular, the basic large deviation principle given in Theorem 1
continues to hold; by virtue of the Gartner-Ellis Theorem \cite{DZ},
it is sufficient that $f_{\epsilon}(\eta)$ is finite and smooth on the
interval $-\infty < \eta < \epsilon^{-1}$.  We omit the analysis that
justifies this extension of the theory already developed.

\vspace{.2in}
\noindent {\bf 6.2 Computed results.} 
To solve the variational principle for the microcanonical model
derived in Theorem 3, we implement the iterative algorithm developed
in \cite{TW} and extended in \cite{DMT}.  Specifically, for given
admissible values $E$ and $\Gamma$ of the energy and circulation
constraints, respectively, we compute the equilibrium macrostate
$\qbar$ that solves
\[
\mbox{ minimize } \; I_{\epsilon}(q) \;\;\;\;
\mbox{ subject to} \;\; H(q)=E, \; C(q)=\Gamma. 
\]
>From an initial guess $q^0$ having the given constaint values
$(E,\Gamma)$, this algorithm defines a sequence $q^k$ of
approximations that converges to a solution $\qbar$ as $k \rightarrow
\infty$.  At each iteration, a variational subproblem defined by
linearizing the energy constraint is solved; its solution, $q^k$, then
satisfies $H(q^k) \ge E, \; C(q^k) = \Gamma$, and $I(q^k) \le
I(q^{k-1})$.  These properties of the iteration step guarantee that
the algorithm is globally convergent \cite{TW}.  In the limit as $k
\rightarrow \infty$, the equality constraint on energy is retrieved,
and the iterative multipliers, $\beta^k$ and $\gamma^k$, which are
determined along with $q^k$, converge to the multipliers $\beta$ and
$\gamma$ associated with $\qbar$.  Experience with this algorithm in a
wide range of statistical equilibrium problems has shown it to be an
efficient and robust method.

We now turn to a description of the computed results for this specific
microcanonical equilibrium problem. 

We compute the equilibrium states $\qbar=\qbar(x_2; E,\Gamma)$ over
the range of constraint values, $0<E\le0.1, -2\le\Gamma\le2$, for both
(a) $r=\infty$ and (b) $r=0.2$.  In each case, we tabulate the
microcanonical entropy $S(E,\Gamma)=-I(\qbar)$.  In Figure 1, we
exhibit the admissible set $\A$ and the concavity set $\C$ for these
two values of $r$.  We recall from Section 4 that $\A$ is the set of
all pairs $(E,\Gamma)$ for which there exists some macrostate $q$
realizing those constraint values $(E,\Gamma)$, and that $\C$ is the
subset of all pairs $(E,\Gamma)$ at which $S$ has a supporting plane.
In Figure 1, $\C$ is indicated by ``equivalence'' and $\A \backslash
\C$ by ``nonequivalence.''  The remarkable result contained in Figure
1 is that, for both $r=\infty$ and $r=0.2$, the concavity set $\C$ is
a relatively small subset of the admissible set $\A$.  In fact, for
any fixed circulation $\Gamma$, the pair $(E,\Gamma)$ lies in $\C$
only for a limited range of energies near the smallest admissible
energy.  For all the energies outside this range, the tangent plane to
$S$ at $(E,\Gamma)$ is not a supporting plane for $S$.  Consequently,
for this range of larger energies, the equivalence of ensembles breaks
down, meaning that the canonical model omits all these microcanonical
equilibrium states.

Another graphical depiction of the nonconcavity present in
$S(E,\Gamma)$ is given in Figures 2 and 3.  In Figure 2 the section of
$S$ versus $E$ at the fixed value $\Gamma=0$ is plotted.  This
entropy-energy curve shows that the inverse temperature $\beta=\d S /
\d E$ is positive only for a small range of low energies below
$E=0.01$.  Throughout the negative temperature range, the entropy
function is slightly concave with respect to $E$, becoming
asymptotically linear for high energy values.  By contrast, the
section of $S$ versus $\Gamma$ at $E=0.05$ plotted in Figure 3 shows
that the entropy-circulation curve is strongly nonconcave for a wide
range of circulation values around zero.  This result suggests that in
this particular problem the nonequivalence of ensembles is largely a
consequence of the circulation constraint.

Figures 1, 2 and 3 also indicate the dependence of the solutions on
the radius of deformation $r$. The nonequivalence set $\A \backslash
\C$ broadens noticably as $r$ is decreased from $r= \infty$ to
$r=0.2$.  Also, the asymmetry in the entropy-circulation curve, which
is a consequence of the skewness $2\epsilon$ of the prior
distribution, increases with decreasing $r$.  These two results
suggest that the effect of the nonlinearity, as measured by
$\epsilon$, is strengthened by a small deformation radius.  From this
behavior we conclude that the breakdown of the equivalence of
ensembles is exacerbated by a weak vertical stratification, which
results in a small $r$.  This conclusion is especially interesting in
the application to the Jovian atmosphere, where large-scale mean flows
such as the permanent zonal winds typically span several radii of
deformation \cite{Marcus}.

Finally, in Figure 4 we display the mean velocity fields associated
with some representative microcanonical equilibrium states.
Specifically, we fix $r=0.2$ and $E=0.05$, and we choose three
representative values of the circulation: (a) $ \Gamma=-0.5$; (b) $
\Gamma=1.4$; (c) $\Gamma=2.0$.  Flow (a) lies within the
nonequivalance set, near the local minimum of $S$ with respect to
$\Gamma$; flow (b) lies near the equivalence-nonequivalence boundary,
which itself is near the global maximum of $S$ with respect to
$\Gamma$; flow (c) lies in the equivalence set.  We draw particular
attention to flow (a), which closely resembles the mean zonal winds
observed in a zone-belt domain of the Jovian atmosphere.  Indeed, this
shear flow consists of a strong westward jet that resides between two
strong eastward jets.  Furthermore, even though the prior
distribution has a positive (cyclonic) skewness, this intense
triple-jet flow has a negative (anticyclonic) circulation.
Interestingly, most of the coherent structures observed on Jupiter and
the other giant planets are anticyclonic.  These general properties on
flow (a), which is representative of the most probable flows in the
nonequivalence set, are not shared by flows (b) and (c).  Instead,
each of these flows consists of one broad westward jet and one narrow
eastward jet, and the total circulation of each of them is positive
(cyclonic).  These weaker shear flow structures are typical of the
equivalence set.

Perhaps our most significant result is that the most probable flows
corresponding to a constraint pairs $(E.\Gamma)$ in the nonequivalence
set are nonlinearly stable, even though they typically fail to satisfy
the well-known stability conditions.  The computed flows discussed above
illustrate this general result quite vividly.  The
most probable flows (a) and (b) displayed in Figure 4 have negative
$\beta$ and fail the often-quoted sufficient condition
\begin{equation} \label{arnold2}
0 \, < \, \frac{d \qbar}{d \psibar} \, < \, \lambda_1 
\end{equation} 
for the second Arnold stability theorem.  Indeed, our computations
show that for the triple-jet flow (a), $d \qbar/d \psibar$ ranges from
$27$ and $78$, while for the qualitatively different flow (b), $d
\qbar/d \psibar$ ranges from $26$ to $42$.  Since $\lambda_1 = \pi^2 +
r^{-2} \approx 35$, we conclude that flow (a), which lies within the
nonequivalence set, is far from satisfying (\ref{arnold2}), while flow
(b), which lies near the equivalence-nonequivalence boundary, comes
closer to fulfilling (\ref{arnold2}).

By contrast, flow (c) in Figure 4, which is a positive temperature
macrostate lying in the equivalence set, satisfies the Rayleigh
condition
\begin{equation} \label{arnold1}
\frac{d \qbar}{d \psibar} \, < \, 0 \, , 
\end{equation}
which implies the first Arnold stability theorem.  In fact, for flow
(c), $d \qbar/d \psibar$ is approximately equal to the constant $-5$
over the domain.

Let us comment further on this gap in the classical stability
criteria.  First, any microcanonical equilibrium $\qbar$, which
corresponds to a constraint pair $(E,\Gamma)$ belonging to the
equivalence set $\C$, is a global minimizer of the associated
information functional $I_{\beta,\gamma}$.  Thus, in principle, the
classical Arnold stability criterion applies, assuming only that the
minimizer is nondegenerate.  Nevertheless, it is possible that the
explicit sufficient condition (\ref{arnold2}) may be violated, even
though $I_{\beta,\gamma}$ is a Lyapunov functional for $\qbar$.
Second, in the case of an equilibrium $\qbar$ which lies slightly
outside the equivalence set, it is possible that the
$I_{\beta,\gamma}$ is a Lyapunov functional, if the microcanonical
entropy $S(E,\Gamma)$ is locally concave at $\qbar$; then, the tangent
plane corresponding to $(\beta,\gamma)$ is locally a supporting plane
for $S$, even though it does not support $S$ globally.  Typically, the
sufficient condition (\ref{arnold2}) is too crude in such a delicate
case. Third, for an equilibrium $\qbar$ which lies far outside the
equivalence set, $I_{\beta,\gamma}$ is not definite at $\qbar$, and
the classical Lyapunov argument based on this functional fails.  Of
course, in this nonequivalent case the sufficient condition
(\ref{arnold2}) is violated.

The above analysis of the various cases possible in the classical
stability criteria notwithstanding, Theorem 7 guarantees that the
microcanonical equilibrium states corresponding to {\em all}
admissible pairs $(E,\Gamma)$ define nonlinearly stable flows,
provided only that a technical nondegeneracy condition is fulfilled.
Given this refined stability result, which makes use of the penalized
Lyapunov functional $L_{\sigma,\tau}^{E,\Gamma}$, it is not necessary
to impose a restrictive condition such as (\ref{arnold2}) to obtain
the stability of most probable flows.  Conversely, it is incorrect to
assume that a steady flow that strongly violates the well-known Arnold
conditions is unstable.  In essence, these conditions are derived by
utilizing a linear combination of two independent conserved quantities
(the energy and a certain enstrophy), while the conservation of each
of these quantities separately constraints the evolution of
perturbations and leads to more refined stability conditions.

%==========================================================================

\section{Acknowledgements}  

\noindent
The authors thank Michael Kiessling and Andrew Majda for helpful
conversations that stimulated aspects of this work.    

%==========================================================================

\newpage
\noindent
{ \large \bf Figure captions} 

\vspace{.2in} \noindent {\bf Fig. 1.}  Admissible set $\A$ and
concavity set $\C$ for the microcanonical variational principle for a
range of constraint values on energy ($0.01 \le E \le 0.1$) and
circulation ($-2 \le \Gamma \le 2$).  The computed boundary of the
admissible set is the dashed curve; the computed boundary of the
concavity, or equivalence, set is the solid curve.  For each
admissible constraint pair $(E,\Gamma)$ in a grid over this range with
$\Delta E=0.0025$ and $\Delta \Gamma = 0.025$, the corresponding
equilibrium macrostate $\qbar$, multipliers $\beta$ and $\gamma$, and
microcanonical entropy $S$ are computed.  A pair $(E,\Gamma)$ is
accepted for the concavity set if the tangent plane at $\qbar$ with
slopes $\beta$ and $\gamma$ lies above the function $S$ throughout the
admissible set.  This computation is displayed for two different
choices of deformation radius: (a) $ r=\infty $ and (b) $ r=0.2$.

\vspace{.1in} \noindent {\bf Fig. 2.}  The section $S(E,0)$ of the
microcanonical entropy for the same variational problem as in Figure
1.  The solid curve is for (a) $r=\infty$, and the dashed curve is for
(b) $r=0.2$.

\vspace{.1in} \noindent {\bf Fig. 3.}  The section $S(0.05,\Gamma)$ of the
microcanonical entropy for the same variational problem as in Figure
1.  The solid curve is for (a) $r=\infty$, and the dashed curve is for
(b) $r=0.2$.  

\vspace{.1in} \noindent {\bf Fig. 4.} Mean velocity fields of the zonal
shear flows determined by the most probable macrostates for the
microcanonical model with $r=0.2$ and $E=0.05$.  Flows corresponding
to three different circulations are displayed: (a) $\Gamma=-0.5$, which
lies within the nonequivalence set; (b) $\Gamma=1.4$, which lies near
the equivalence-nonequivalence boundary; (c) $\Gamma=2.0$, which
lies in the equivalence set.

\newpage

\epsfig{file=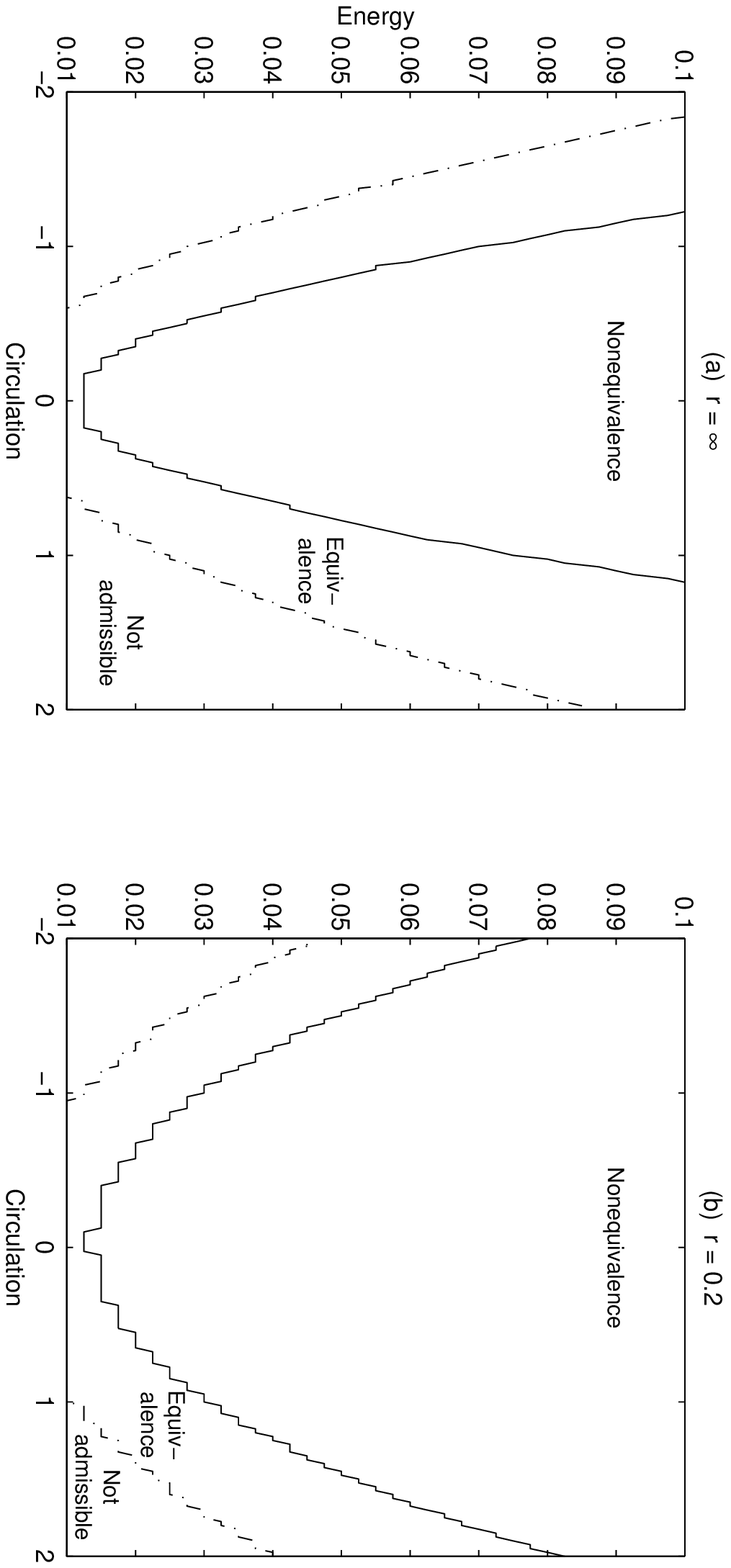}

\newpage

\epsfig{file=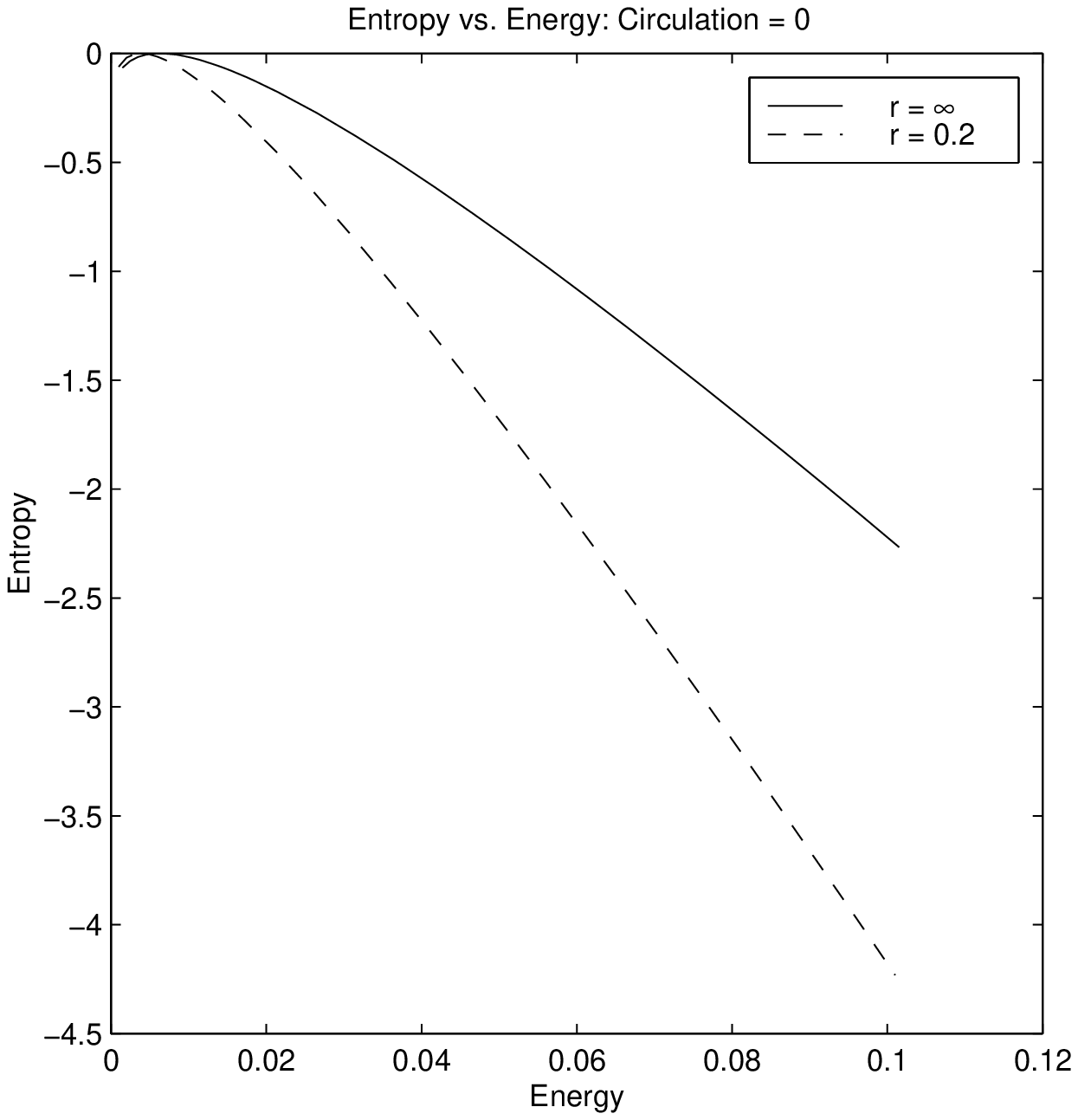}

\newpage

\epsfig{file=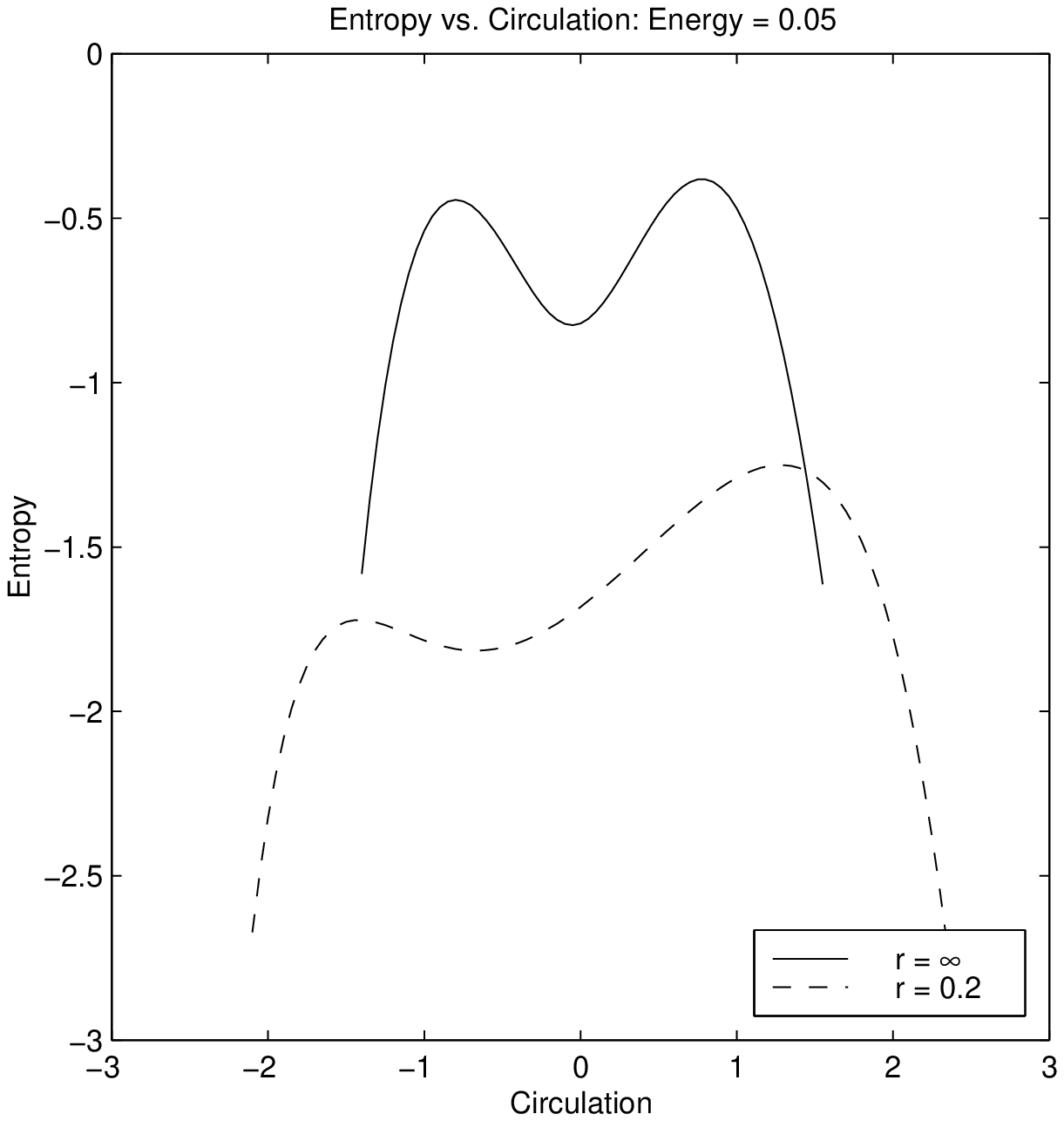}

\newpage

\epsfig{file=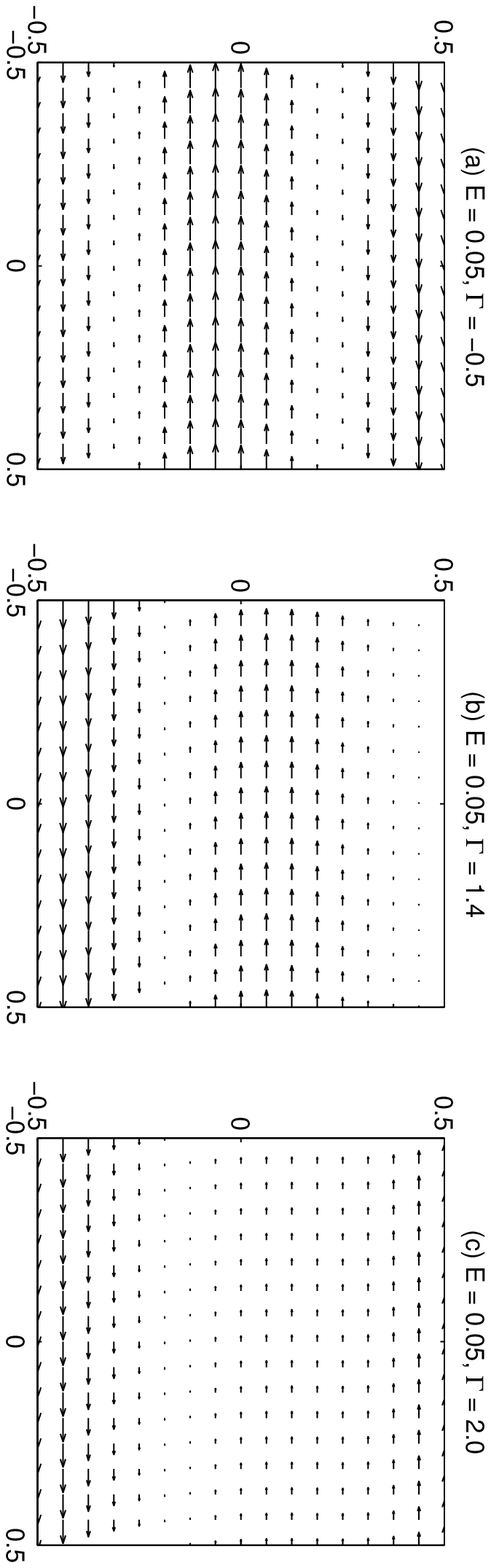}

\end{document}